# Phonon properties and unconventional heat transfer in quasi-2D $Bi_2O_2Se$ crystal

Jan Zich[1,2], Antonín Sojka[1], Petr Levinský[2], Martin Míšek[2], Kyo-Hoon Ahn[2], Jiří Navrátil[1], Jiří Hejtmánek[2], Karel Knížek[2], Václav Holý[3,4], Dmitry Nuzhnyy[5], Fedir Borodavka[5], Stanislav Kamba[5] and Čestmír Drašar[1,*]

[1]*University of Pardubice, Faculty of Chemical Technology, Studentská 573, 53210 Pardubice, Czech Republic*
[2]*Institute of Physics of the Czech Academy of Sciences, Cukrovarnická 10/112, 162 00 Prague 6, Czech Republic*
[3]*Department of Condensed Matter Physics, Faculty of Mathematics and Physics, Charles University, Ke Karlovu 3, 121 16 Praha 2, Czech Republic*
[4]*Masaryk University, Department of Condensed Matter Physics, Kotlářská 2, 61137 Brno, Czech Republic*
[5]*Institute of Physics of the Czech Academy of Sciences, Na Slovance 2, 182 00 Prague 8, Czech Republic*

**ABSTRACT**. $Bi_2O_2Se$ belongs to a group of quasi-2D semiconductors that can replace silicon in future high-speed/low-power electronics. However, the correlation between crystal/band structure and other physical properties still eludes understanding: carrier mobility increases non-intuitively with carrier concentration; the observed $T^2$ temperature dependence of resistivity lacks explanation. Moreover, a very high relative out-of-plane permittivity of about 150 has been reported in the literature. A proper explanation for such a high permittivity is still lacking. We have performed infrared (IR) reflectivity and Raman scattering experiments on a large perfect single crystal with defined mosaicity, carrier concentration and mobility. Five of the eight phonons allowed by factor group theory have been observed and their symmetries determined. The IR spectra show that the permittivity measured in the tetragonal plane is as high as $\varepsilon_r \approx 500$, and this high value is due to a strong polar phonon with a low frequency of ~34 $cm^{-1}$ (~1 THz). Such an unusually high permittivity allows the screening of charge defects, leading to the observation of high electron mobility at low temperatures. It also allows effective modulation doping providing a platform for high performance 2D electronics. DFT calculations suggest the existence of a very low frequency acoustic phonon ~14 $cm^{-1}$ (~0.4 THz). Both the low frequency phonons cause anomalous phonon DOS, which is reflected in the unconventional temperature dependence of the heat capacity, $c_M \approx T^{3.5}$. The temperature-dependent, two-component group velocity is proposed to explains the unusual temperature dependence of the thermal conductivity, $\kappa \approx T^{1.5}$.

## I. INTRODUCTION.

The search for high-speed, low-power electronic materials has largely focused on 2D materials. Quasi-2D $Bi_2O_2Se$ (Figure 1) is one of the most promising materials in this regard [1–4], outperforming both silicon-based electronics and other 2D materials in many respects [5]. This material exhibits extraordinarily high charge carrier mobility [6,7], and allows the formation of an intrinsic insulating layer ($Bi_2SeO_5$) in an epitaxial manner, thus facilitating easy formation of gates in transistors. In addition, 2D $Bi_2SeO_5$ even appears to enhance electron mobility in adjacent $Bi_2O_2Se$ single crystals, although the reason is not clear [8]. In general, understanding the relationship between structure/defect and physical properties is still a matter of debate. Its metallicity is attributed to a metal-insulator transition (MIT) with a very low critical concentration $n_c \approx 10^{15} cm^{-3}$, mainly due to a very high relative permittivity, $\varepsilon_r \approx 150$ [9] or to self-modulation doping due to native defects (Se-vacancies - $V_{Se}$) located in the conduction band, causing their spontaneous ionization at any temperature [10]. The sporadic observation of a hidden direct band gap in this case underlines the complexity of this material [11]. At 2 K, the mobility of the carriers reaches values up to 470,000 or 280,000 $cm^2V^{-1}s^{-1}$ for different authors [7,12]. Given an average effective carrier mass $m_{ef} \cong 0.14 \ m_e$, such mobilities can be regarded as ultrahigh, yielding long relaxation times of $\approx 3 \cdot 10^{-12}$ s and $\approx 5 \cdot 10^{-13}$ s, respectively. Note that these values of mobility exceed theoretical predictions ($\approx 100,000 \ cm^2V^{-1}s^{-1}$, for a carrier concentration of $n=10^{16} \ cm^{-3}$) [13]. Even the high permittivity mentioned above ($\varepsilon_r \approx 150$) still lacks an explanation [9]. The deviation of the stoichiometry (Bi/Se=1.4 -2.3) in apparently perfect $Bi_2O_2Se$ single crystals is worrisome [14]. Again, this suggests a possible large concentration of native defects or rather the presence of "invisible" foreign phases (Bi-Se or $Bi_2SeO_5$ based). The "unzipped" layers reported in ref [15] may also play an important role in this regard. It is also interesting to note that the mobility of the carriers generally increases with their concentration [5,9,16]. In addition, the reported effective electron masses range from $0.03m_e$ [14] to $0.23m_e$ [16], which is not a convincing result.

*Contact author: Cestmir.Drasar@upce.cz

Our measurements show that the in-plane permittivity is more than three times higher than reported out-of-plane permittivity, resulting in a critical MIT concentration, $n_c \approx 2 \times 10^{14}$ cm$^{-3}$. This is a much lower number compared to $n_c \approx 1 \times 10^{16}$ cm$^{-3}$ from ref [9]. Moreover, we provide clear evidence that low-frequency optical phonon with very small damping and high dielectric strength is the cause of such high permittivity. Interestingly, our experimental frequency of the lowest optical phonon ($\nu = 34$ cm$^{-1}$) is lower than the theoretical values ($\nu = 55$-$60$ cm$^{-1}$) [17,18]. However, our DFT calculation obtained the phonon with a frequency as low as 24 cm$^{-1}$. Such a discrepancy between theory and experiment, combined with the large deviation in stoichiometry (Bi/Se=1.4 - 2.3), suggests a link between the defect structure and the permittivity of Bi$_2$O$_2$Se. The low energy optical and acoustic phonons cause a strong deviation of the phonon density of states (PDOS) from the theoretical $\omega^2$ dependence in the low frequency range. This leads to an unconventional temperature dependence of the heat capacity and the thermal conductivity. Although the electrical conductivity of the sample is high, we did not observe any plasma from free carriers in the IR and THz region. The plasma frequency must be in the microwave range due to the high mobility of the charge carriers and high permittivity. As a result, the sample becomes transparent in the mid-IR region. Furthermore, in contrast to previous reports, we detected all four theoretically predicted Raman modes in our samples and compared their frequencies with theoretical values.

The data reported in the literature come from various samples prepared by CVT and PVT or Bridgman methods, and intentionally or unintentionally doped. The lateral dimensions of the samples range from dozens of micrometers to a small number of mm and thicknesses from a few nm to 500 mm. This may probably be the reason for the inconsistency of some published data. For example, low-temperature mobility ranges from $10^2$-$10^3$ [14] to $10^5$ cm$^2$V$^{-1}$s$^{-1}$ [7,12]. It should be emphasized here that we used structurally well-characterized, large, thick and apparently perfect single crystals with large domains and low mosaicity (0.7°) for our measurements. The present paper sheds more light on open questions (e.g., various mobilities in different samples) in the literature and significantly deepens the understanding of this exciting material.

*Contact author: Cestmir.Drasar@upce.cz

## II. EXPERIMENTAL

### A. Crystal growth

Bi$_2$O$_2$Se single crystals were prepared from bismuth chunks (SigmaAldrich, 5N) ground in an MM500 nano oscillating mill at 30 Hz for 20 min, selenium shards (SigmaAldrich, 5N) ground in an agate mortar, and Bi$_2$O$_3$ powder (AlfaAesar, 5N) calcined (decarbonized) at 450 °C for 30 min in an argon flow and subsequently rapidly cooled. These powders were mixed in a stoichiometric ratio in an agate mortar and sealed in a quartz ampoule under vacuum <10$^{-3}$ Pa. Ampoules were rotated in furnace at 400 °C for 10 days at 1 rpm to ensure homogenous solid-state reaction to Bi$_2$O$_2$Se. For crystal growth, ampoules were placed in a temperature gradient of 850/800 °C and the powder was distributed over the entire length of the ampoules. The growth of the crystals was carried out for 2 weeks, followed by controlled cooling for 1 weeks to 150 °C to reach near equilibrium. The aim was to achieve a "defect-free" structure with increased chemical and mechanical stability of the samples. The largest single crystal was about 7 mm long, 5 mm wide and 0.3 mm thick (Figure 1). Although Raman spectra show that the crystal probably grows with the help of Bi$_2$SeO$_5$ (see Figure S14 in Supplemental Materials (SM) [19], other data below suggest that it nevertheless grows in Se-rich conditions (see SM section 8 for explanation). Specifically, high resolution X-ray diffraction (HRXRD) analysis in our samples suggests the formation of Se$_{Bi}$ antisites and the electron concentration and mobility are relatively low, both indicating Se-rich conditions (see sections 1 and 8 in (SM) [19]). We note that it is difficult to determine whether chemical or physical vapor transport predominates.

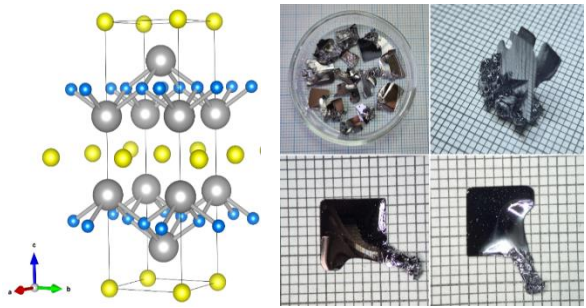

FIG 1. (left) Crystal structure of tetragonal Bi$_2$O$_2$Se. (right) Photographs of typical single crystals. We selected the best plane-parallel samples for physical measurements. The gray, blue and yellow spheres represent Bi, O and Se atoms, respectively.

### B. High resolution XRD) – structural analysis

The aim of the high-resolution X-ray diffraction (HRXRD) experiments was to confirm the overall crystallographic quality and reveal the domain structure of the single crystals and to provide the most accurate lattice parameters and lattice position occupancy for a given crystal. Measurements were performed on a Rigaku Smartlab X-ray diffractometer equipped with a 9kW CuKα rotating anode (45 kV/200 mA). An HRXRD setup with a parabolic multilayer mirror and a 2x220Ge channel-cut monochromator on the primary side, and a one-dimensional detector was used to measure long symmetric 2Θ/ω scans and reciprocal space maps. Low-resolution overview reciprocal space maps were measured in the microdiffraction setup using a two-dimensional detector to search for foreign phases.

### C. Transport measurements

All transport properties were measured using the PPMS apparatus (Quantum Design, Ltd.), temperature range 2-300 K, and magnetic field up to 14 T. Rectangular samples of approximately 2×5×0.5 mm$^3$ were characterized using the Electrical Transport Option in a 4-wire configuration for electrical resistivity and Hall effect measurements (with magnetic field orientation along the crystallographic c-axis, i.e. with in-plane electric fields consistent with optical measurements). The Thermal Transport Option was used to determine the thermal conductivity and heat capacity option was used to determine the heat capacity.

### D. IR spectroscopy

Low-temperature IR reflectivity spectra were measured at near-normal incidence using the Fourier transform IR spectrometer Bruker IFS 113v equipped with Si bolometer detector operating at 1.6 K in the range of 20-650 cm$^{-1}$ (0.6–20 THz). Continuous-flow He cryostats (Optistat, Oxford) with polyethylene windows was used for the low-temperature IR experiment. Room-temperature spectra were obtained in the range up to 3000 cm$^{-1}$ (90 THz) using pyroelectric detectors. An FTIR Nicolet Nexus spectrometer was used for room-temperature transmission and reflection measurements up to 10,000 cm$^{-1}$. The IR spectra were fitted using a generalized four-parameter oscillator model with the factorized form of the complex permittivity:

$$\varepsilon(\omega) = \varepsilon_1(\omega) + i\varepsilon_2(\omega) = \varepsilon_\infty \prod_j \frac{\omega_{LOj}^2 - \omega^2 + i\omega\gamma_{LOj}}{\omega_{TOj}^2 - \omega^2 + i\omega\gamma_{TOj}} \quad (1)$$

where $\omega_{TOj}$ and $\omega_{LOj}$ stand for transverse and longitudinal frequencies of the *j-th* polar phonon, respectively, and $\gamma_{TOj}$ and $\gamma_{LOj}$ mark the corresponding damping constants [20]. This model is suitable for the spectra analysis with large TO-LO splitting. The high-frequency permittivity, $\varepsilon_\infty$, originating from electronic absorption processes, was obtained from the frequency-independent reflectivity tail at room temperature above the phonon frequency. The complex permittivity, $\varepsilon(\omega)$, is related to the reflectivity $R(\omega)$ by the relation:

$$R(\omega) = \left|\frac{\sqrt{\varepsilon(\omega)}-1}{\sqrt{\varepsilon(\omega)}+1}\right|^2. \quad (2)$$

### E. Raman spectroscopy

The polarized Raman spectra were recorded in a back-scattering geometry using a RM1000 Renishaw micro-Raman spectrometer with the 514.5 nm line of an Ar$^+$ ion laser at a power of about 20 mW (~ 200 μW on the sample), in the spectral range 10–700 cm$^{-1}$. The spectrometer was equipped with Bragg grating filters enabling good stray light rejection. Low-temperature Raman spectra were acquired using a LINKAM temperature cell THMS350V. The diameter of the laser spot on the sample surface was ∼5μm, and the spectral resolution was better than 1.5 cm$^{-1}$.

### F. DFT calculations

Density functional theory (DFT) calculations were used to calculate the electronic structures. The calculations of defect energies were performed in the $3 \times 3 \times 1$ supercell using the program WIEN2k [21]. The program uses the full potential linearized augmented plane-wave (FP LAPW) method with the dual basis set. In the LAPW methods, the space is divided into atomic spheres and the interstitial region. The electron states are then classified as the core states fully contained in the atomic spheres, and the valence states. The radii of the atomic spheres were taken 2.325 a.u. for Bi, 1.86 a.u. for O and 2.325 a.u. for Se. The positions of the atoms in the whole supercell were optimized individually for each type of defect. The number of *k*-points in the irreducible part of the Brillouin zone was 172 (7×7×7), which is sufficient for the large unit cell used.

The basic electronic structure for the post-processing phonons was obtained using the VASP package [22,23]. We used the projector-augmented wave (PAW) potentials [24] and the generalized gradient approximation (GGA) [25] for the exchange-correlation functional. We performed a structure optimization that resulted in a fully relaxed parameter set of a = 3.9264 Å, c = 12.4039 Å, and z$_{Bi}$ = 0.3529 for the *I4/mmm* (No. 139) lattice with a *k*-mesh of

*Contact author: Cestmir.Drasar@upce.cz

16×16×4 points in the BZ for the rectangular 2 formula units (f.u.) cell and a plane-wave cutoff of 600 eV. Phononic properties were calculated using the Phonopy code [26,27] based on the finite difference approach (FDA). The expanded cell of 4×4×1 including 32 f.u. was used to take into account for an appropriate number of atomic displacements. A fine q-mesh of $100^3$ for the 1st BZ was used to capture the detailed structure of the phononic density of states. The dielectric properties were investigated using density functional perturbation theory (DFPT) implemented in VASP, with the rectangular 2 f.u. cell and a k-mesh of 16×16×4. The electronic contribution of the static permittivity $\varepsilon_{el} \equiv \varepsilon_\infty$ and Born effective charges Z* for each atomic site were carried out in terms of 3×3 tensors to evaluate the effects of LO-TO splitting in the phononic structure. The ionic contribution of $\varepsilon_{ion}$ was derived with two different methods of DFPT and FDA showing a good agreement between them.

### III. RESULTS AND DISCUSSIONS

#### A. XRD – structural analysis

Three $Bi_2O_2Se$ structures are available in the literature in refs [18,28,29]; the analysis of our HRXRD data is more consistent with the first one. Also, the works [28,29] seem to contain typos in the Wyckoff positions – the correct positions are Bi/4e;O/4d;Se/2a; see section B. This correction makes all the structures identical. Thus, we consider the structure [18] with tetragonal space group $I4/mmm$. Figure 2 shows the large HRXRD 2Θ/ω scans of the investigated samples. The sharp maxima correspond to the reciprocal lattice points of 00L, where L is an even number. The measured scans were fit to a semi-kinematic diffraction model [30] and the distances of the (001) basal planes were determined from the fits. We observe a nice agreement between the measured, theoretical and fitted diffraction maxima. Refinement of the HRXRD data allows us to calculate the concentration of native point defects. Theoretically, Se-vacancies ($V_{Se}$) and selenium atoms in Bi positions ($Se_{Bi}$ antisites) are the most abundant native defects; see SM, Table S1 [19] and refs [5,31–33]. Note that the energies of defect formation are otherwise largely inconsistent in the literature. Our experiments rather suggest the presence of $Se_{Bi}$ defects, since we obtained $V_{Se}\cong 0$ (Figure S1) and the fit was slightly worse than that assuming only $Se_{Bi}$ defects; see section 1 of SM for details. This indicates Se-rich growth conditions that are consistent with the experiment, i.e. the reaction of $Bi_2O_3$ with the quartz ampoule [34], and refinement suggests a crystal composition of approximately $Bi_{1.9}O_2Se_{1.1}$. With respect to the EDS elemental analysis (SM, section 1) the non-stoichiometry is likely to be much smaller ($Bi_{1.997}O_{1.998}Se_{1.005}$) than predicted by HRXRD. On the other hand, the literature reports Bi/Se ratio of 1.4 to 2.3 for formally flawless crystals suggesting rich point defect chemistry. Therefore, the slight shift in stoichiometry ratios can induce changes in defect chemistry with a significant effect on charge transport [14].This conclusion is difficult to prove by other methods. See discussion in SM, section 1.

We measured reciprocal spatial maps around the reciprocal lattice points 00 6 and 00 10 to assess the mosaicity patterns from spot extensions in Debye rings (Figure S2 in SM [19]). We concluded that the lateral size of the mosaic blocks is larger than the X-ray coherence length (≈1 $\mu m$) and the angular mosaicity is approximately 0.7 deg. We used the well-established Williamson-Hall plot to estimate the vertical size (thickness) of the mosaic blocks D (Figure S3 in SM [19]); the average value is D ≅370 nm. We then used the Cohen-Wagner plot to derive the lattice parameter c≅1.2207 nm (Figure S3 in SM [19]), which is comparable to the published values of 1.2213 nm or 1.22164 nm in refs [28,29], and a lattice parameter $a$≅0.3886 nm (Figure S4 in SM [19]). All these data indicate an excellent quality of the single crystals. However, we observe two very weak reflections of 2Θ≈9° and 18° (the red arrows in Figure 2 and Figure S6 in SM [19]), which are difficult to assign to any known phase; see discussion in SM [19], Figures S5 and S6.

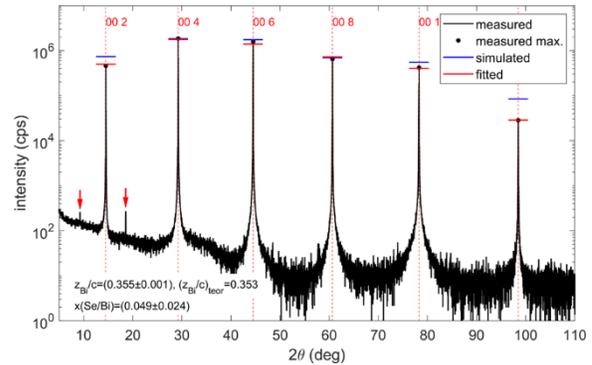

FIG 2. HRXRD symmetric 2Θ/ω scans of one of the samples examined. The theoretical diffraction maxima 00(2n) are indicated by the red vertical dotted lines. Short horizontal blue and red lines mark the peak intensities determined from the published structure [29] and from the fit, respectively, the black dots highlight the measured intensity maxima. $z_{Bi}/c$ is the relative interplanar distance between the Bi atomic layers, and its value in ref [29] is indicated as well. x($Se_{Bi}$) denotes the fraction of the Bi sites occupied by Se atoms, obtained from refinement. x($Se_{Bi}$) =0.049 corresponds approximately to the formula $Bi_{1.9}O_2Se_{1.1}$. The same procedure shows that the fraction of selenium vacancies $V_{Se}$ is close to zero in our samples (see section 3.2 for details).

*Contact author: Cestmir.Drasar@upce.cz

Two weak reflections, 2Θ≈9° and 18°, denoted by red arrows remain unexplained. They can be attributed to monoclinic Se (2Θ≈9°) and $Bi_2SeO_5$ (2Θ≈18°). For details, see SM [19], Figures S5 and S6.

### B. Transport measurements

Given the peculiarities of $Bi_2O_2Se$ mentioned in the introduction, it is desirable to measure its transport properties. Figure 3 shows the temperature dependence of the Hall concentration and mobility. Metallic conductivity with electrons as the majority carriers at a relatively low concentration (≈5 x $10^{16}$ cm$^{-3}$) is consistent with Se-rich growth conditions leading to the formation of predominantly $Se_{Bi}$ antisites (see SM [19], Figures S7 and S8). The carrier concentration can be used as an indicator of Se rich/Se poor samples with ≈1 x $10^{17}$ cm$^{-3}$ as a borderline [9]. However, the electron concentration can be attributed to $V_{Se}$ rather than to $Se_{Bi}$. In particular, the former forms metallic states in the conduction band (CB), while electrons from the latter must be excited (compare SM [19], Figures S8 and S9). The background carrier concentration corresponds to one $V_{Se}$ per ≈270 000 Se atoms, but the slight increase in carrier concentration above 100 K may be related to the thermal excitation of carriers from $Se_{Bi}$. In contrast to the Hall concentration $n_H$, the Hall mobility of the electrons drops sharply from $\mu_H \approx$ 30 000 cm$^2$V$^{-1}$s$^{-1}$ at ≈ 30 K to $\mu_H \approx$ 200 cm$^2$V$^{-1}$s$^{-1}$ at T=300K due to the scattering on gradually populated phonons. Below ≈ 30 K, we observe an increase in mobility with temperature, consistent with the dominance of scattering on ionized impurities. Qualitatively, the same temperature dependence was predicted theoretically in ref [13]. Due to the relatively low carrier concentration/mobility, we were unable to measure the SdH effect. Note that a higher electron concentration would be associated with plasma resonance in the IR region (we do not observe plasma resonance in this sample), which would preclude the analysis of low-frequency phonons discussed below; see section 8 of the SM [19] for details.

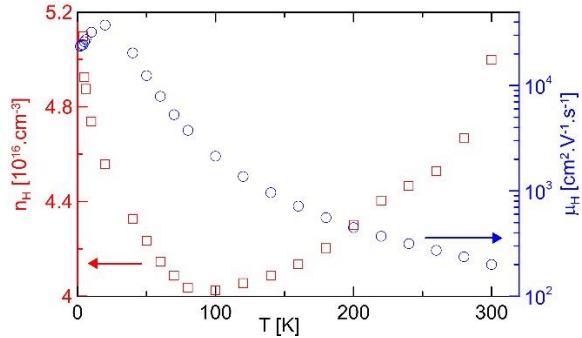

FIG 3. Temperature dependence of Hall concentration $n_H$ and mobility $\mu_H$ for the $Bi_2O_2Se$ crystal used for IR and Raman spectroscopy.

### C. Phonons

From the theoretical factor group analysis of the $Bi_2O_2Se$ structure with $D_{4h}$ point group symmetry with five atoms in the primitive cell, the phonons at Γ point of the Brillouin zone have the following symmetries [35,36]:

$$\Gamma_{acoustic} = A_{2u} + E_u$$
$$\Gamma_{optical} = A_{1g}(x^2 + y^2, z^2) + 2A_{2u}(z) + B_{1g}(x^2 - y^2) + 2E_g(xz, yz) + 2E_u(x, y). \quad (3)$$

The modes $E_u$ and $E_g$ are double degenerate. Among the optical modes, $A_{1g} + 2E_g + B_{1g}$ represent 4 Raman active modes, while the other $2A_{2u} + 2E_u$ represent four modes active in $E \| c$ and $E \| a, b$ polarized IR spectra, respectively. Eigenvectors of all optical phonons are shown in [35,36]. If the O atoms were located at position 4c ($D_{2h}$ symmetry reported in Ref. [28]) instead of correct 4d ($D_{2d}$ symmetry), the factor-group analysis of the optical phonons would be quite different:

$$\Gamma_{optical} = A_{1g}(x^2 + y^2, z^2) + 2A_{2u}(z) + E_g(xz, yz) + 3E_u(x, y) + B_{2u}(-). \quad (4)$$

We will show below that this analysis does not match our spectra because it predicts only $3E_u$ IR active phonons in the tetragonal plane (we see 2). It means, our spectra confirm the *I4/mmm* crystal structure with the following atom positions Bi/4e; O/4d; Se/2a.

#### 1. IR reflectivity

The broadband IR spectra of both the transmission and reflectivity at 300 K are shown in Figure S10 in SM [19]. Figure 4 shows the far IR reflectivity measurement of the $Bi_2O_2Se$ single crystal as a function of temperature.

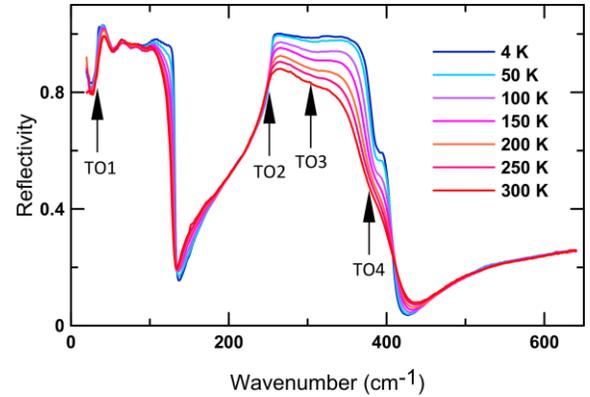

FIG 4. Reflectivity of $Bi_2O_2Se$ single crystal in far IR range. Two distinct transversal optical modes TO1 and TO2 with $E_u$ symmetry can be seen. We observe a negligible softening of the modes with temperature, but significant drop in damping with decreasing temperature. The oscillations close to TO1 are due to multiphoton absorption and limited measurement accuracy below 100 cm$^{-1}$. For details, see discussion in section D.

*Contact author: Cestmir.Drasar@upce.cz

Although only two broad reflection bands can be clearly observed, we used four oscillators for the fitting (SM [19], Figures S11, S12) to achieve a good fitting agreement with the experiment. The four oscillators apparently match the theoretical prediction in Eq. (3). However, for symmetry reasons, only two $E_u$ symmetry modes should be observed in our $E \perp c$ polarized IR spectra. The frequencies of these two modes are labeled TO1 and TO2. The remaining two modes (labeled TO3 and TO4) cannot be $2A_{2u}$ modes from symmetry reasons and can be caused either by multiphonon absorption or possibly by phonons from the $Bi_2SeO_5$ ultrathin layer that could theoretically grow on the surface of the $Bi_2O_2Se$ crystal. Therefore, we prepared a pure $Bi_2SeO_5$ crystal, the spectrum of which is shown in (SM [19], Figure S13). This crystal has a very rich phonon spectrum, but none of its phonons match the TO3 and TO4 frequencies in the $Bi_2SeO_5$ spectrum. A multiphonon origin of these modes is therefore more likely. Note that $\Delta\varepsilon$ of TO3 and TO4 modes are negligibly small. The results of the fits are summarized in Table I and the calculated complex permittivity from the obtained phonon parameters is shown in Figure 5. Note that the low frequency phonon near 34 cm$^{-1}$ has $\Delta\varepsilon$ close to 500. This is a value comparable to the strength of ferroelectric soft modes. This phonon alone raises the static permittivity above 500. It is the low frequency and large $\Delta\varepsilon$ of this TO1 mode that makes the crystal lattice unstable and susceptible to ferroelectric phase transition under tensile strain, as theoretically predicted in ref [13].

TABLE I. Fitting parameters of IR reflectivity spectra using the four-parameter model. The high frequency permittivity, $\varepsilon_\infty$=13.3. All four oscillators can be compared with the theoretical and experimental values shown in Table II, but the weak TO3 and TO4 modes are most likely of multiphonon origin, since $2A_{2u}$ symmetry modes cannot be detected in our $E \perp c$ polarized spectra.

| Temperature | Number | $\omega_{TO}$ (cm$^{-1}$) | $\gamma_{TO}$ (cm$^{-1}$) | $\omega_{LO}$ (cm$^{-1}$) | $\Gamma_{LO}$ (cm$^{-1}$) | $\Delta\varepsilon$ |
|---|---|---|---|---|---|---|
| **4 K** | TO1 | 33.7 | 2.3 | 132.8 | 1.1 | 496.0 |
| | TO2 | 253.3 | 1.0 | 299.2 | 33.9 | 15.0 |
| | TO3 | 300.0 | 33.7 | 383.1 | 16.9 | 0.16 |
| | TO4 | 386.1 | 21.4 | 407.8 | 6.9 | 0.04 |
| **300 K** | TO1 | 34.3 | 7.8 | 129.8 | 5.5 | 451.0 |
| | TO2 | 250.3 | 14.5 | 293.3 | 65.4 | 13.6 |
| | TO3 | 300.6 | 70.3 | 375.3 | 47.9 | 1.4 |
| | TO4 | 386.1 | 59.0 | 418.6 | 40.4 | 0.2 |

The experimental TO1 frequency is much lower than theoretical values in refs [18,36], reflecting some uncertainty in the DFT calculations. Although unlikely, this can also be attributed to defects in the crystal structure. In particular, the HRXRD data suggest a high concentration of selenium Se$_{Bi}$ defects which could potentially cause such a redshift (SM, section 1). Figure 5 shows that the mode with the lowest energy and damping is responsible for the high "static" in-plane permittivity of $\varepsilon_r \approx 510$. This value is considerably higher than that reported in the existing literature for the c-axis $\varepsilon_r < 200$ [9], which is indicative of the anisotropic nature of the material. Note that such a high permittivity combined with a small effective electron mass creates favorable conditions for modulation doping. The concept of modulation doping is mainly faced with the problem of limited charge carrier release due to electrostatic charging of the modulation phase [37–39]. This limitation logically decreases significantly with increasing permittivity of the host matrix. In addition, the fact that the mobility in Bi$_2$O$_2$Se increases with electron concentration strongly suggests that this type of doping is inherently active in this material [9,16]. In particular, the disconnected ("unzipped") layers [15] correspond to fluctuations in the work function [14]. Both findings suggest the presence of a 2D defect of polyatomic size. Given the HRXRD data, the most likely candidates would be 2D complex defects composed of a large number of Se$_{Bi}$ (or with respect to DFT calculations, V$_{Se}$). On the other hand, so-called self-modulation doping due to V$_{Se}$ located in the conduction band [10] is a less likely alternative, since it is inconsistent with the fact that the mobility in Bi$_2$O$_2$Se increases with electron concentration [9].

*Contact author: Cestmir.Drasar@upce.cz

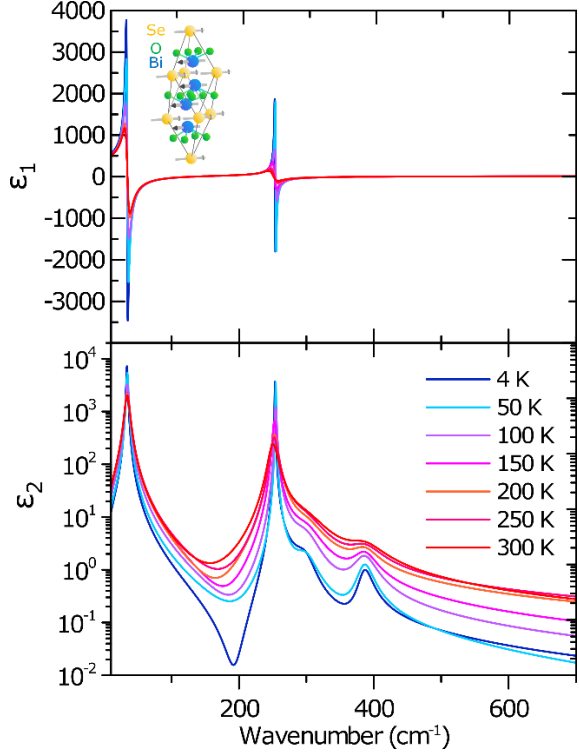

FIG 5. Real ($\varepsilon_1$) and imaginary ($\varepsilon_2$) parts of the complex permittivity obtained by fitting the reflectivity spectra. The strong resonance of the optical phonon at $\nu \cong 34$ cm$^{-1}$ is largely responsible for the high static permittivity. The imaginary part of the permittivity indicates four oscillators in accordance with the loss function (SM [19], Figure S11). The inset shows the displacement of atoms of the low frequency $E_u$ mode.

### 2. Raman spectroscopy

The Raman spectra measured on the largest sample are shown in Figure 6. The individual symmetries revealed are consistent with the geometry of the experiments. We observe four Raman modes at $\approx$ 58, 165, 363 and 435 cm$^{-1}$ (Table II). The latter three are in good agreement with the theoretical values $A_{1g}$, $B_{1g}$, $E_g$ in refs [18,35,36]. The first $E_g$ mode seen at 58 cm$^{-1}$ has lower energy than predicted theoretically, but this may be due to antisite defects Se$_{Bi}$ or vacancies V$_{Se}$ in the crystal lattice, as in the case of the polar $E_u$ mode. An attempt has been made in the literature [17,36,40] to explain the mode seen in the Raman spectrum at 55 cm$^{-1}$ as a defect-induced polar $E_u$ mode, but our IR spectra clearly show that the frequency of the $E_u$ phonon is 20 cm$^{-1}$ lower. The broad wings of the first $E_g$ mode can be explained by modes from the secondary phase of Bi$_2$SeO$_5$ (SM [19], Figures S14 and S15). Bi$_2$SeO$_5$ seems to be involved in the growth of the crystal and occurs naturally at growth interfaces [34]. The spectra were measured in four different polarization configurations: $x(yz)\bar{x}$, $x(zz)\bar{x}$, $z(xx)\bar{z}$, and $z(yx)\bar{z}$. The orientation of the $a$ and $b$ axes was not exactly known, so we mark the polarization vectors in the tetragonal plane as the $x$ and $y$ axes. The $c$-axis is perpendicular to the crystal surface, so it is clearly defined and is parallel to $z$. To collect the spectra in x($zz$)$\bar{x}$, and x($yz$)$\bar{x}$, configurations and to avoid the influence of the secondary phase of Bi$_2$SeO$_5$, which can grow on the edge of the crystal, the sample was divided into two parts and measured on the fresh edge. Table II shows a comparison of the theoretical and experimental Raman frequencies. It is very surprising that all four phonons are observed in the $zz$ spectrum, although only one $A_{1g}$ mode should be seen in this spectrum. The spectrum was measured at the broken edge of the crystal and since the crystal was difficult to break due to the strong 2D bonds (it had to be bent several times), it is possible that the crystal orientation was changed at the broken edge, or the crystal symmetry was locally broken. The presence of $E_g$ mode at 58 cm$^{-1}$ in x($zz$)$\bar{x}$, spectrum can be the result of polarization leakage, since this is the strongest one in the x($yz$)$\bar{x}$, spectrum.

TABLE II. Calculated and measured frequencies (in cm$^{-1}$) of IR and Raman modes in Bi$_2$O$_2$Se. IR-active modes are shown in red rows, and Raman-active modes in the blue ones. The TO3 and TO4 modes seen in Figure 4 near 300 and 386 cm$^{-1}$ are close to $E_u$ and $A_{2u}$ symmetry modes but are most likely of multiphonon origin. In ref [17], the ~55 cm$^{-1}$ phonon is derived from Raman spectra and assigned as defect-induced IR mode. [a] data from [36], [b] data from [18], [c] data from [35], [d] data from [17]

| Symmetry | Activity | Xu[a] Calculated | Pereira[b] Calculated | Cheng[c] Calculated | Kim[d] Measured | This work Measured |
|---|---|---|---|---|---|---|
| $E_u$ | IR | 54.8 | 59.2 | | ~55 | 33.8 (at 4 K) |
| $A_{2u}$ | IR | 65.0 | 64.5 | | | |
| $E_g$ | Raman | 67.3 | 72.0 | 67.99 | 78 | 57.8 (at 80 K) |
| $A_{1g}$ | Raman | 162.9 | 165.7 | 159.89 | 158.8 | 165.3 (at 80 K) |
| $E_u$ | IR | 268.0 | 293.9 | | | 253.3 (at 4 K) |
| $B_{1g}$ | Raman | 354.3 | 369.4 | 364.02 | 360 | 362.7 (at 80 K) |
| $A_{2u}$ | IR | 377.8 | 402.8 | | | |
| $E_g$ | Raman | 433.3 | 444.0 | 428.68 | 434 | 435.3 (at 80 K) |

*Contact author: Cestmir.Drasar@upce.cz

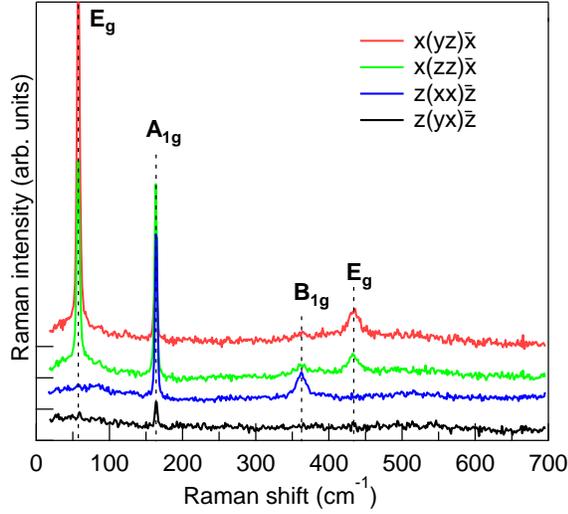

FIG 6. Polarized Raman spectra of $Bi_2O_2Se$ single crystal measured at 80 K in four different polarization configurations.

### D. Heat capacity and heat conductivity at low temperatures

Figure 7 shows the heat capacity $C_p$ and the thermal conductivity $\kappa$ of $Bi_2O_2Se$. Both experimental quantities show unconventional behavior. In particular, both quantities do not follow the expected $T^3$ law. With respect to the IR data above, this can be attributed to the presence of a low frequency optical phonon with unusually strong dispersion near the center of the Brillouin zone (and hence the group velocity). To confirm this conclusion, we have performed DFT calculations.

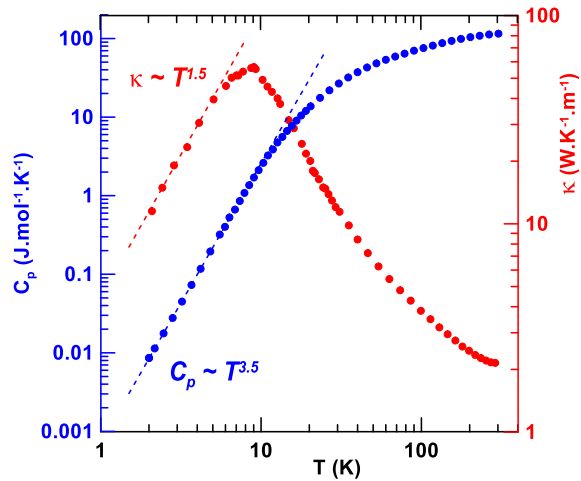

FIG 7. Heat capacity, (blue solid circles) and thermal conductivity (red solid circles) versus temperature on a logarithmic scale. Both do not follow the $T^3$ law in the low temperature region. This peculiarity is related to the low frequency acoustic and optical modes (Figure 8).

*Contact author: Cestmir.Drasar@upce.cz

The theoretical phonon band structure and the total and partial DOS are shown in Figure 8. Including the effect of the TO-LO splitting in the calculation resulted in a significant frequency difference between the longitudinal optical (LO) phonons and their corresponding transverse optical (TO) modes. The calculation indicates a very low energy TO phonon $\approx 25$ cm$^{-1}$ ($\approx 3$ meV), which is even lower than our experimental value of the transverse optical mode TO1 $\approx 34$ cm$^{-1}$. This transversal mode can alternatively be attributed to the theoretically predicted IR-active $E_u$ mode $\nu = 55$ cm$^{-1}$ implying a significant redshift of the experimental eigenfrequency [36]. We do not have a simple explanation for this discrepancy at the moment, but we have observed the same resonance frequency for other samples with slightly different composition, so this low experimental frequency (compared to the theoretical one) is unlikely to be caused by defects in the crystal lattice. It can also be seen that the acoustic mode has a very low frequency of $\approx 14$ cm$^{-1}$ ($\approx 1.8$ meV) in the M-point of the Brillouin zone. Both low frequency modes appear as cusps in the phonon DOS, see the inset in Figure 8. These phonons contribute to the unconventional temperature dependence of the heat capacity ($c_p \sim T^{3.5}$).

Similar values of the low energy acoustic phonon in the M point (labeled Z in some papers) of 14 cm$^{-1}$ and optical phonons of 20 cm$^{-1}$ have been calculated ref. [35], while other authors report higher values around 20 and between 35-55 cm$^{-1}$, respectively refs. [40–43]. The calculated high-frequency permittivity $\varepsilon_\infty = 13.9$ and static permittivity $\varepsilon_r \sim 620$ in the $xy$-plane (see SM [19], Table S2 for details) are in a good agreement with our experimentally determined $\varepsilon_\infty = 13.3$ and 510, respectively. The calculated static permittivity along the $z$-direction, $\varepsilon_r \sim 135$ is in a good agreement with the out-of-plane permittivity of about 150 reported in ref. [9] and it is due to the higher frequency of the $A_{2u}$ phonon. The behavior of the phonon DOS deviates strongly from the theoretical $\omega^2$ dependence. Thus, a simple Debye model fit of the temperature dependent heat capacity over the entire temperature range is in very poor agreement with the experimental data (Figure 9). In contrast, using calculated phononic DOS gives an excellent agreement (Figure 10). The steep increase in phonon DOS due to the two low energy phonon modes (Figure 8) mimics a compound with a rather low Debye temperature $\Theta_D$ (as if it had a low Young's modulus and heavy atoms). With increasing temperature, the higher energy part of the phonon DOS is populated and $\Theta_D$ increases (Figure 9). Such a structure can therefore be considered as a "thermal composite" of two different structures, hard (Bi-O-Bi) and soft (Bi-Se-Bi), with significantly different elastic properties. In this view, the thermal communication between these two systems is limited at low temperatures due to a large energy gap in the phononic energy spectrum. In particular, almost all three-phonon events taking place in the low-energy Bi-Se-Bi layers are unable to involve any phonon from the high-energy Bi-O-Bi layers [44] (Figure 56, pp. 130-145), red (Se/Bi) and blue (O) areas in Figure 8.

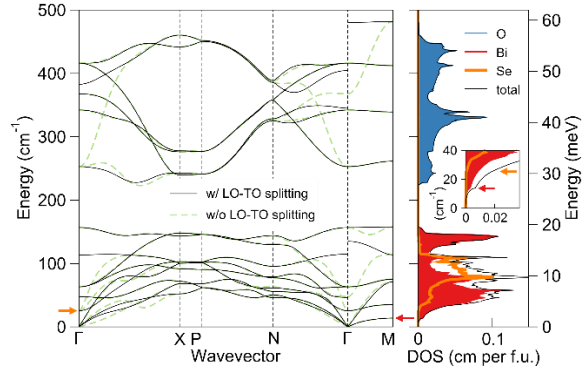

FIG 8. (left) Phononic band structure of $Bi_2O_2Se$. Low frequency acoustic phonon (14 cm$^{-1}$) at the M point, and optical phonon (25 cm$^{-1}$) at the Γ point, are marked by arrows. The corresponding Brillouin zone with special k-vector points is displayed in SM [19], Figure S16. (right), Total and partial phononic DOS of $Bi_2O_2Se$. The two cusps due to low-frequency acoustic and optical phonons are marked by arrows in the inset with the low energy detail. The effect of defects is not incorporated.

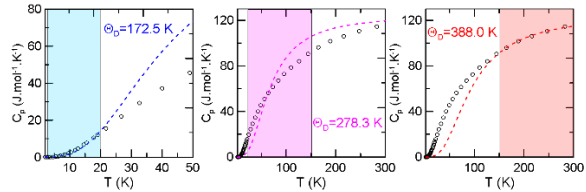

FIG 9. Heat capacity of $Bi_2O_2Se$ as a function of temperature. The figure compares the theoretical Debye model (dashed lines) with experimental data (black circles). The fitted temperature ranges are shown in color. The Debye temperature $\Theta_D$ is strongly temperature dependent, mainly due "two component" nature of this compound [43]. This is because of the peculiar energy dependence of the phonon DOS Figure 8 (SM [19], Figure S17).

To support the idea of a thermal composite, we calculated the temperature dependent group velocity $v_g$ for the red (Bi-O-Bi) and blue (Bi-Se-Bi) layers in Figure 8 (for details see SM, section 6). The result is shown in Figure 11. The two velocities are very different below 300K, and so are the acoustic refractive indices [44]. This results in frequent reflections of phonons on both sides of the interface, implying the reduced heat exchange between the layers. Moreover, the heat transport by oxygen-based phonons is limited to a region above T≈40K. In the low temperature region, only Se/Bi-based phonons transport heat. The steep drop in their group velocity explains the lower exponent of the temperature dependent thermal conductivity ($\kappa \sim T^{1.5}$).

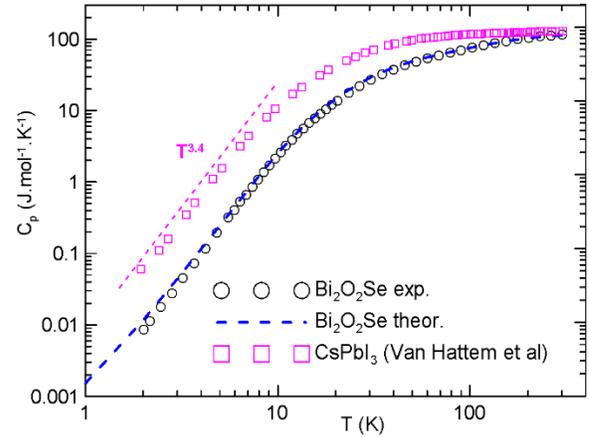

FIG 10. Calculated temperature dependent heat capacity compared with the experimental data. The experimental data are almost identical with the data published in ref. [43]. A similar temperature dependence is also shown for $CsPbI_3$ in ref. [45].

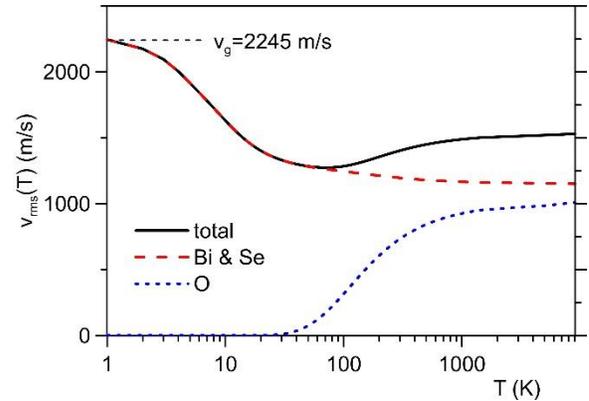

FIG 11. Calculated temperature dependent group velocities (their rms values, $v_{rms}$) for the two DOS regions red (Se/Bi) and blue (O) areas in Figure 8. Thermal communication between the corresponding layers Bi-Se-Bi and Bi-O-Bi is limited due to the large difference in group velocities, which largely causes reflection of phonons. The situation is analogous to the principle of optical fibers at low temperatures This picture also justifies the use of combination of Debye and Einstein models in Figure S17c. For details see SM, section 6. Low temperature limit of the group velocity (rms), $v_g$=2245 ms$^{-1}$.

This view inspires the idea of treating the system as two formally independent structures. It can be composed of two 2D structures or, alternatively, of a 3D, low-energy, Debye-like structure and a second, high-energy, Einstein-like structure, see SM [19], Figure S17. This is consistent with the two-component model [43]. In addition, such a composite structure exhibits the presence of low-frequency, highly dispersive optical phonons, which contribute to the thermal conductivity and heat capacity. All this explains the formally strong temperature dependence of $\Theta_D$ in this compound. It does not mean that the structure as a whole is of 2D nature and that heat and charge transport must be predominantly 2D [43]. Although, the 2D nature of charge transport was reported for Se-poor samples in contrast to Se-rich ones in [14].

*Contact author: Cestmir.Drasar@upce.cz

## IV. CONCLUSIONS

We have carried out a study of $Bi_2O_2Se$ single crystals to demonstrate and explain some of its special properties. Specifically, these are the very high permittivity and the unconventional temperature dependence of the heat capacity and thermal conductivity. In contrast to previous works, we have shown experimentally and theoretically that this material possesses two very low frequency phonons: one acoustic at ≈14 cm$^{-1}$ (≈1.8 meV) and one optical near ≈24 cm$^{-1}$ (≈3 meV). While the optical mode induces a markedly high relative in-plane permittivity ($\varepsilon_r$>500), both modes contribute to the unconventional low-temperature dependence of the thermal capacity and conductivity. They increase the phonon DOS in the low temperature region, which explains the higher order temperature dependence of the heat capacity, $c_p \sim T^{3.5}$. On the other hand, the steep decrease of the phonon group velocity with temperature explains the lower exponent of the temperature dependent thermal conductivity ($\kappa \sim T^{1.5}$).

Our data and theoretical models provide additional clues to explain the high mobility. Not only the high permittivity but also the type of defects seems to have an effect on the electron mobility. Using HRXRD, we demonstrated that Se-rich growth conditions led to the formation of Se$_{Bi}$ antisites rather than Se vacancies, suggesting that the high carrier mobility is related to both the presence of a low-frequency optical phonon and the type of native point defect. Consistent with the literature, Se vacancies, V$_{Se}$ appear to be weaker scattering centers compared to Se$_{Bi}$. Unfortunately, it is more difficult to grow comparably sized and quality single crystals under Se-poor conditions to demonstrate this. Significantly, the high permittivity of this material provides ideal conditions for modulation doping. This suggests that the larger 2D polyatomic defects that can naturally occur in the $Bi_2O_2Se$ structure may also serve as doping centers, explaining the ultrahigh mobility of some samples reported in the literature. On the other hand, it shows that once a suitable "doping phase" is found, it will be possible to use modulational doping of $Bi_2O_2Se$ in artificial 2D heterostructures to significantly improve its transport properties.

## ACKNOWLEDGMENTS

This work was supported by the Czech Science Foundation (Project No. 22-05919S). This work was also supported by the project TERAFIT - CZ.02.01.01/00/22_008/0004594 co-financed by the European Union and the Ministry of Education, Youth and Sports of the Czech Republic, and by the grant of the Ministry of Education, Youth and Sports of the Czech Republic (grant LM2023037). We want to thank Marketa Jarošová for the EDS measurements.

*Contact author: Cestmir.Drasar@upce.cz

*Contact author: Cestmir.Drasar@upce.cz

*Contact author: Cestmir.Drasar@upce.cz

# Phonon Properties and Anomalous Heat Transfer in Quasi-2D $Bi_2O_2Se$ Crystal

*Supplemental Material*

Jan Zich[1], Antonín Sojka[1], Petr Levinský[2], Martin Míšek[2], Kyo-Hoon Ahn[2], Jiří Navrátil[1], Jiří Hejtmánek[2], Karel Knížek[2], Václav Holý[3,4], Dmitry Nuzhnyy[5], Fedir Borodavka[5], Stanislav Kamba[5] and Čestmír Drašar[1,*]

[1]University of Pardubice, Faculty of Chemical Technology, Studentská 573. 53210 Pardubice, Czech Republic

[2] Institute of Physics of the Czech Academy of Sciences, Cukrovarnická 10/112, 162 00 Prague 6, Czech Republic

[3]Department of Condensed Matter Physics. Faculty of Mathematics and Physics, Charles University, Ke Karlovu 3. 121 16 Praha 2, Czech Republic

[4] Masaryk University, Department of Condensed Matter Physics and CEITEC, Kotlářská 2, 61137 Brno, Czech Republic

[5]Institute of Physics of the Czech Academy of Sciences, Na Slovance 2, 182 00 Prague 8, Czech Republic

[*]Corresponding author: cestmir.drasar@upce.cz

## 1. Structural analysis – HRXRD experiments. EDS

We used the high-resolution X-ray diffraction (HRXRD) experiments to assess the structural quality of single crystals. In this document we provide additional information to that presented in the main text. Figure S1 shows the large HRXRD 2Θ/ω scan of the investigated sample. The sharp maxima correspond to the reciprocal lattice points of 00L. where L is an even number. The measured scans were fit to a semi-kinematic diffraction model [1] and the distances of the (001) basal planes were determined from the fits. Refinement of the HRXRD data allows us to calculate the concentration of native point defects.

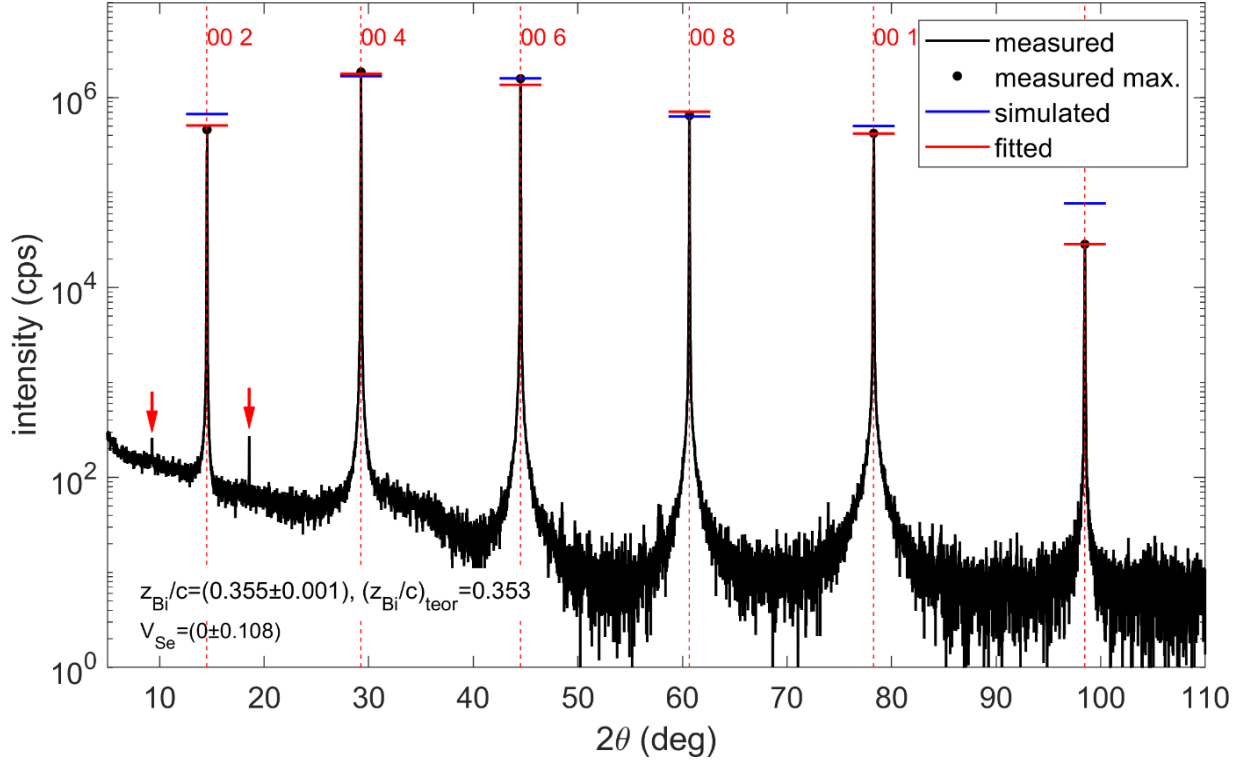

FIG S1. HRXRD symmetric 2Θ/ω scans of $Bi_2O_2Se$ investigated (analogous to Figure 2 in the main text for Se vacancies, $V_{Se}$). The theoretical diffraction maxima 00(2n) are indicated by the red vertical dotted lines. Short horizontal blue and red lines mark the peak intensities determined from the published structure [29] and from the fit, respectively. The black dots highlight the measured intensity maxima. $z_{Bi}/c$ is the relative interplanar distance between the Bi atomic layers and its value in ref. [2] is indicated as well. $x(V_{Se})$ denotes the fraction of the selenium vacancies obtained from refinement. Two weak reflections, 2Θ≈9° and 18°, denoted by red arrows remain unexplained. They can perhaps be attributed to monoclinic Se (2Θ≈9°) and $Bi_2SeO_5$ (2Θ≈18°). For details, see Figures S5 and S6.

Unfortunately, the measured data are not sensitive enough to be able fit the densities of $Se_{Bi}$ and $V_{Se}$ defects simultaneously and we have to assume the presence of one defect type only. The result of the fit setting the $V_{Se}$ density to zero is shown in Figure 2 in the main text; the fitted curve for zero $Se_{Bi}$ density is displayed in Figure S1. We observe a nice agreement between the measured, theoretical and fitted diffraction maxima in both structural models, however the residuum for fitted $Se_{Bi}$ density is 0.076 which is slightly smaller than in the latter model (0.080). In addition to the occupancies we also fitted the interplanar distance of Bi layers, which corresponded quite well with the tabulated value in both fitting models as well as the B-factor, which covers both thermal lattice displacements (expressed by the Debye-Waller factor $D = \exp\left[-B\left(\frac{\sin(\theta)}{\lambda}\right)^2\right]$ ) and static displacements caused by defects. On the other hand, if we set the $Se_{Bi}$ density to zero and fit the $V_{Se}$ density we obtain $V_{Se}\cong0$ and non-physically large B factor of about $5\ \text{Å}^2$. Other defect types like O vacancies or Se-Bi antisite defects did not give any reasonable fits. This is indicative of Se-rich growth conditions.

The value of the $Se_{Bi}$ density from the fit $x(Se_{Bi}) = 0.049$ corresponds approximately to the formula $Bi_{1.9}O_2Se_{1.1}$. However, this value is burdened with an error of nearly 50% so that the value of the defect density can be treated with caution; moreover, other defect types not considered here (stacking faults, dislocations) influence the heights of the diffraction maxima as well. Therefore, the result of the fit only indicates the presence of $Se_{Bi}$ defects.

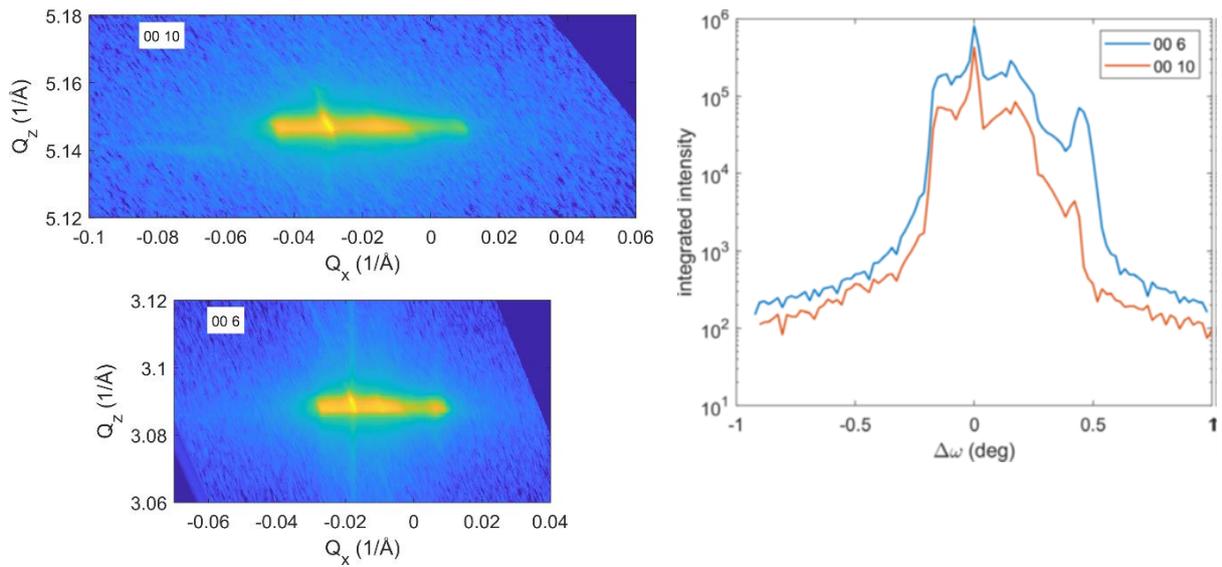

FIG. S2 (left) Reciprocal space maps measured around the reciprocal lattice points 00 6 (bottom) and 00 10 (top) on $Bi_2O_2Se$. The intensities are plotted logarithmically, and the color scale spans over 5 decades. (right) The reciprocal-space maps plotted in left part integrated in angular space in the direction perpendicular to the Debye rings. The horizontal coordinate is the angular deviation $\omega$ from the diffraction maximum and the solid lines depict the intensities around the reciprocal lattice points 00 6 (blue), and 00 10 (red). The fact that the curves for 00 6 and 00 10 are almost identical in angular space indicates that the $\omega$–broadening of the maxima is caused by angular misorientation of the mosaic blocks and not by the finite block size. Thus, the lateral size of the mosaic blocks is larger than the x-ray coherence width ($\approx 1\ \mu m$); the angular mosaicity is roughly 0.7 deg [3].

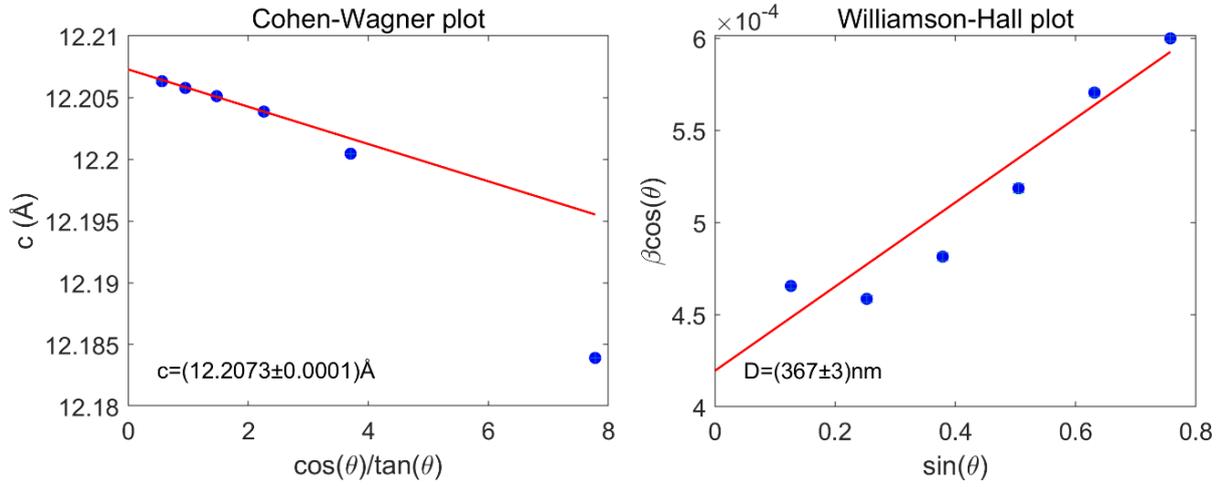

FIG. S3. (left) Cohen-Wagner plot used to derive c-lattice parameter of $Bi_2O_2Se$ single crystal. (right) The Williamson-Hall plot [4] used to determine vertical size (thickness) of coherent domains D in the single crystal. By extrapolating the linear part of this function to $\sin(\Theta) \rightarrow 0$ we estimated the values of $D \cong 370\ nm$.

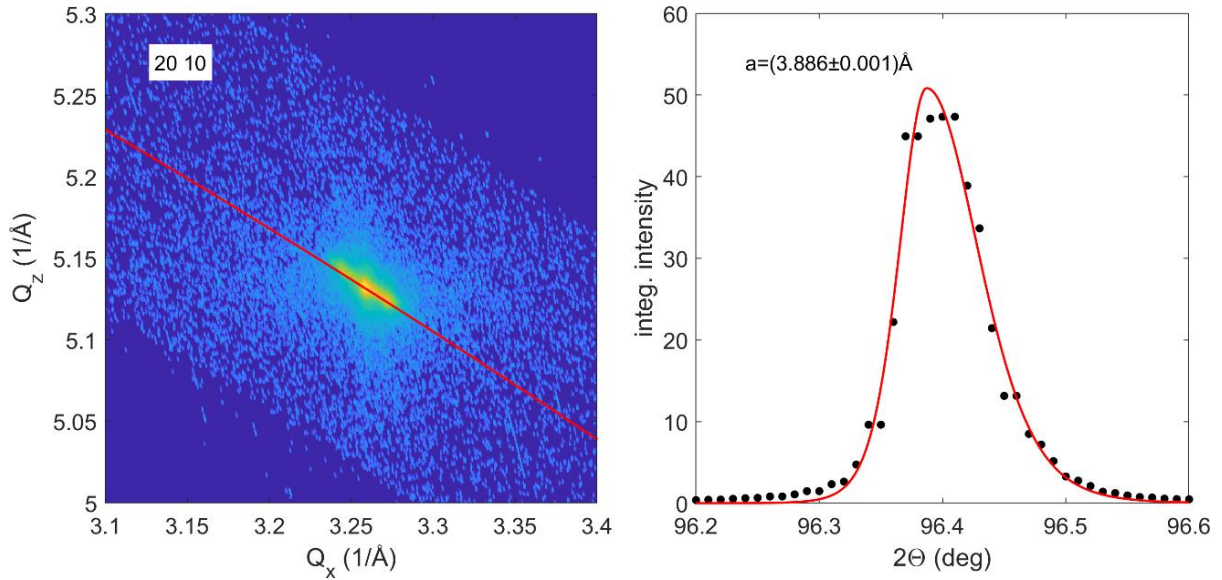

FIG. S4. (left) Coplanar asymmetric reciprocal space map, diffraction 2 0 10 . The red line corresponds to the Debye ring 2Θ=const. From the fact that diffraction maximum is elongated along the Debye ring we conclude that the diffraction broadening is caused solely by angular misorientation ω of the mosaic blocks. In the right panel we plotted the intensity integrated over $\omega$, from which we determined the tetragonal lattice parameter a$\cong 0.3886$ nm.

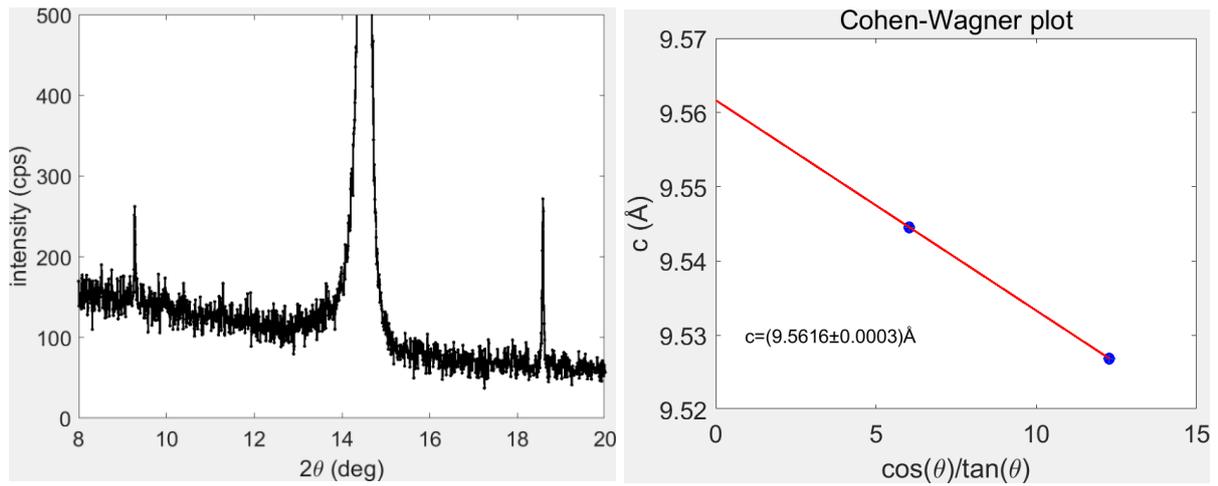

FIG. S5. Experimental analysis of two weak extraordinary reflections (2Θ≈9° and ≈18°) in a HRXRD sample of $Bi_2O_2Se$, assuming that both are associated with the same phase (structure). We obtain the vertical lattice parameter of unknown phase c≅0.956 nm from Cohen-Wagner plot. We failed to attribute both the reflections to single phase from the Bi-Se-O system.

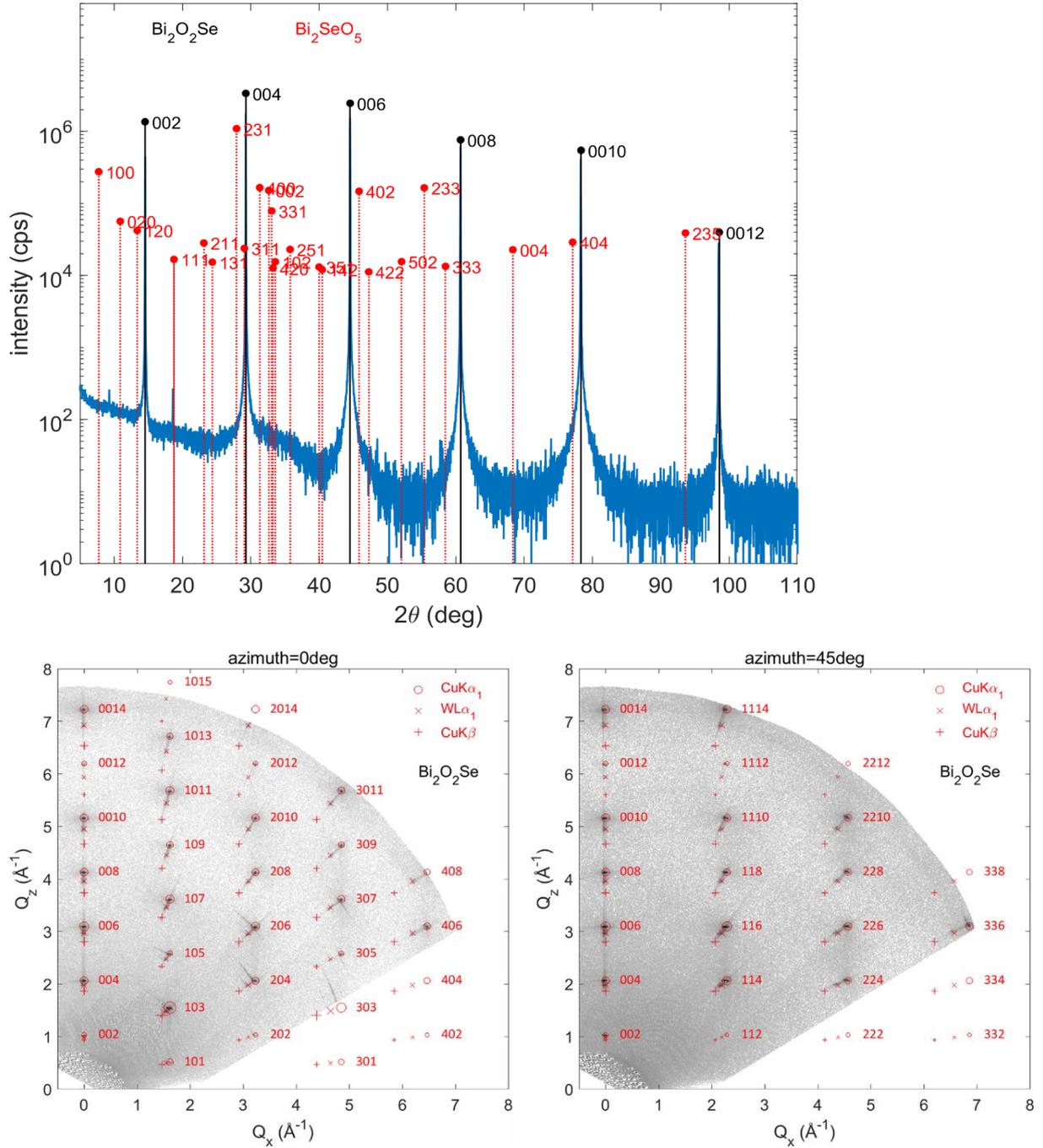

FIG. S6. (top) Symmetric $2\Theta/\omega$ scan of the $Bi_2O_2Se$ sample. The black and red dots and vertical lines denote the theoretical diffraction maxima of the $Bi_2O_2Se$ and $Bi_2SeO_5$ phases, respectively. Obviously, the peak at $2\Theta \approx 18\text{deg}$ can be ascribed to the 111 reflections of $Bi_2SeO_5$ secondary phase, however the nature of the other tiny peak at $2\Theta \approx 9\text{deg}$ is completely unclear. The interplanar distance following from the position of the $2\Theta \approx 18\text{deg}$ peak is $c \cong 0.956$ nm, which to our knowledge does not correspond well to any known phase in the Bi-Se-O system. The Raman spectra of the same $Bi_2O_2Se$ crystal indicate that secondary $Bi_2SeO_5$ phase is present at least at the edges of sample, (see the main text). Thus, they can perhaps be attributed to some of the monoclinic Se allotropes ($2\Theta \approx 9°$) and $Bi_2SeO_5$ ($2\Theta \approx 18°$). (bottom) Low-resolution overview reciprocal space maps used for foreign phase search, taken in two azimuthal directions. The red circles and crosses denote the peak positions calculated for $CuK\alpha_1$, $WL\alpha_1$, and $CuK\beta$ lines, which were present in the x-ray spectrum due to the absence of a monochromator. No foreign phases are detected.

## EDS elemental analysis

Energy dispersive analysis was performed on the sample to determine the true stoichiometry of the crystal under investigation. The results averaged over six spot measurements distributed over the sample area give the stoichiometry $Bi_{1.997}O_{1.998}Se_{1.005}$ (±0.01, 0.02, 0.01, respectively). We see that the non-stoichiometry is much smaller than predicted by HRXRD in the main text. On the other hand, the literature reports Bi/Se ratio variations as large as 1.4 to 2.3 [5] for formally perfect crystals. This suggests that a small shift in the stoichiometry ratios can induce changes in the defect chemistry with a significant effect on charge transport. Although it seems unlikely, we assume the existence of combinations of defects that can preserve the original stoichiometry, e.g. $2V_{Se}+Se_{Bi}+V_O$. For comparison we present the elemental analysis for an attempted Se poor sample, which was not large enough for optical measurements.

| EDS analysis of Se-rich sample used for optical measurements in the main text. | | | | | | |
|---|---|---|---|---|---|---|
| Measurement no. | Bi (at.%) | O (at.%) | Se (at.%) | Corresponding stoichiometry | Bi | O | Se |
| 1 | 39.32 | 20.32 | 40.36 | | 2.02 | 1.97 | 1.02 |
| 2 | 40.97 | 19.71 | 39.32 | | 1.97 | 2.05 | 0.99 |
| 3 | 39.58 | 20.21 | 40.21 | | 2.01 | 1.98 | 1.01 |
| 4 | 39.95 | 20.14 | 39.92 | | 2.00 | 2.00 | 1.01 |
| 5 | 39.67 | 20.22 | 40.1 | | 2.01 | 1.98 | 1.01 |
| 6 | 40.25 | 20 | 39.75 | | 1.99 | 2.01 | 1.00 |
| | | | | Average | 1.997 | 1.998 | 1.005 |
| | | | | Error ± | 0.01 | 0.02 | 0.01 |

| EDS analysis of Se-poor sample presented for comparison. | | | | | | |
|---|---|---|---|---|---|---|
| Measurement no. | Bi (at.%) | O (at.%) | Se (at.%) | Corresponding stoichiometry | Bi | O | Se |
| 1 | 41.34 | 19.5 | 39.16 | | 1.96 | 2.07 | 0.98 |
| 2 | 38.92 | 20.42 | 40.66 | | 2.03 | 1.95 | 1.02 |
| 3 | 38.66 | 20.41 | 40.94 | | 2.05 | 1.93 | 1.02 |
| 4 | 40.49 | 19.92 | 39.59 | | 1.98 | 2.02 | 1.00 |
| 5 | 42.03 | 19.24 | 38.74 | | 1.94 | 2.10 | 0.96 |
| 6 | 41.34 | 19.5 | 39.16 | | 1.99 | 2.01 | 0.99 |
| | | | | Average | 1.991 | 2.014 | 0.995 |
| | | | | Error ± | 0.04 | 0.06 | 0.02 |

## 2. DFT - electronic partial DOS and formation energies of native defects

We used DFT calculations to solve the native point defects. Table S1 summarizes the formation energies of native point defects that can occur in $Bi_2O_2Se$. For comparison, we used two different functionals (potentials), GGA and mBJ. The two types of point defects have reasonably low emergence energies. The Se antisite replacing Bi, $Se_{Bi}$, and the vacancy at the Se site, $V_{Se}$. The former has, on average, the lowest formation energy. The Figures S7 to S9 show the corresponding partial DOS.

We have noticed that DFT calculations are less reliable (especially for defects) for this particular compound [6–8]. With respect to this uncertainty, we assume that the background electron concentration corresponds to $V_{Se}$ (all literature data agree) and the weak activation in Figure 3 corresponds to $Se_{Bi}$, see also comments in point 1. Of course, the activation may also be due to other defects.

**Table S1** Formation energy of native point defects in $Bi_2O_2Se$ for two different functionals (potentials). GGA and mBJ.

| Stoichiometry of supercel 3x3x1 $Bi_2O_2Se$ | Description | Absolute cohesive energy (eV) **GGA** potential | Defect formation energy (eV) | Absolute cohesive energy (eV) **mBJ** potential | Defect formation energy (eV) |
|---|---|---|---|---|---|
| Bi(36)O(36)Se(18) | Stoichiometric | -347.868 | 0.000 | -297.010 | 0.000 |
| Bi(35)O(36)Se(18) | Vacancy Bi | -341.200 | 6.669 | -288.080 | 8.930 |
| Bi(36)O(35)Se(18) | Vacancy O | -341.320 | 6.548 | -292.610 | 4.400 |
| Bi(36)O(36)Se(17) | Vacancy Se | -343.101 | **4.767** | -294.364 | **2.646** |
| Bi(36)O(35)Se(19) | Antisite $Se_O$ | -342.617 | 5.252 | -292.723 | 4.287 |
| Bi(35)O(36)Se(19) | Antisite $Se_{Bi}$ | -344.113 | **3.755** | -293.770 | **3.240** |

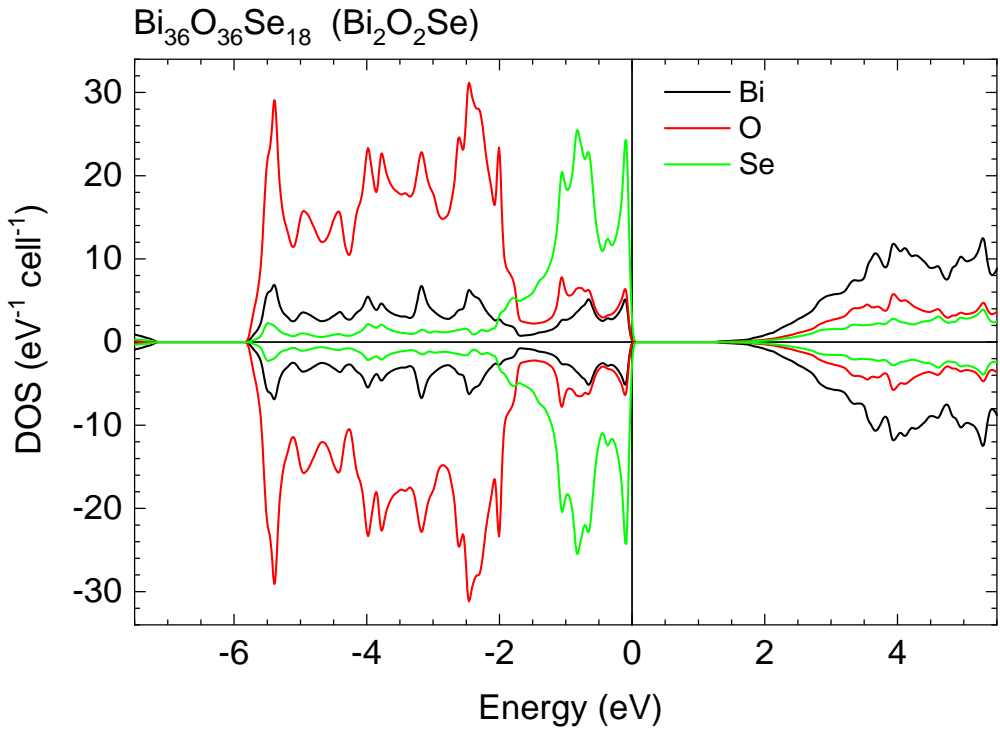

FIG. S7. Partial DOS for defect-free $Bi_2O_2Se$ using mBJ potential. The calculated energy gap $E_{mBJ}$ = 1.2 eV is slightly higher than the experimental optical energy gap $E_{G(opt)} \approx 0.9$ eV (Figure S20). Energy gap obtained by GGA potential $E_{GGA}$ = 0.45 eV is much lower.

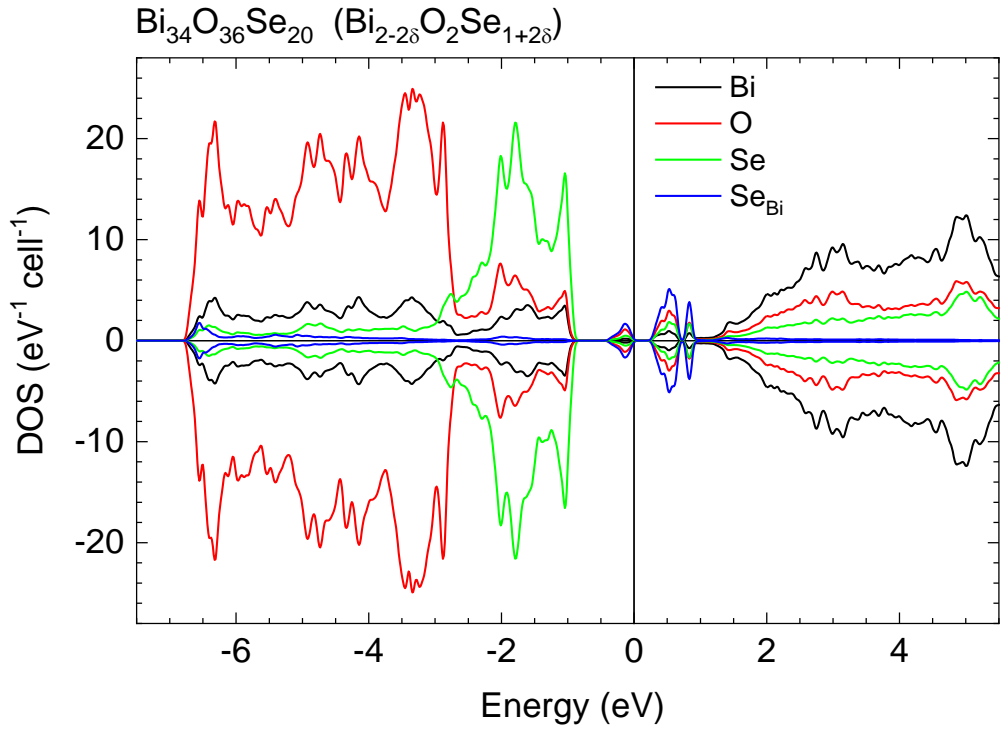

FIG. S8. Partial DOS using mBJ potential for $Bi_2O_2Se$ with antisite defect $Se_{Bi}$. The energy gap is narrower due to formation of states in the middle of the original gap. The black vertical line denotes Fermi energy. It indicates semiconducting to metallic behavior depending on temperature.

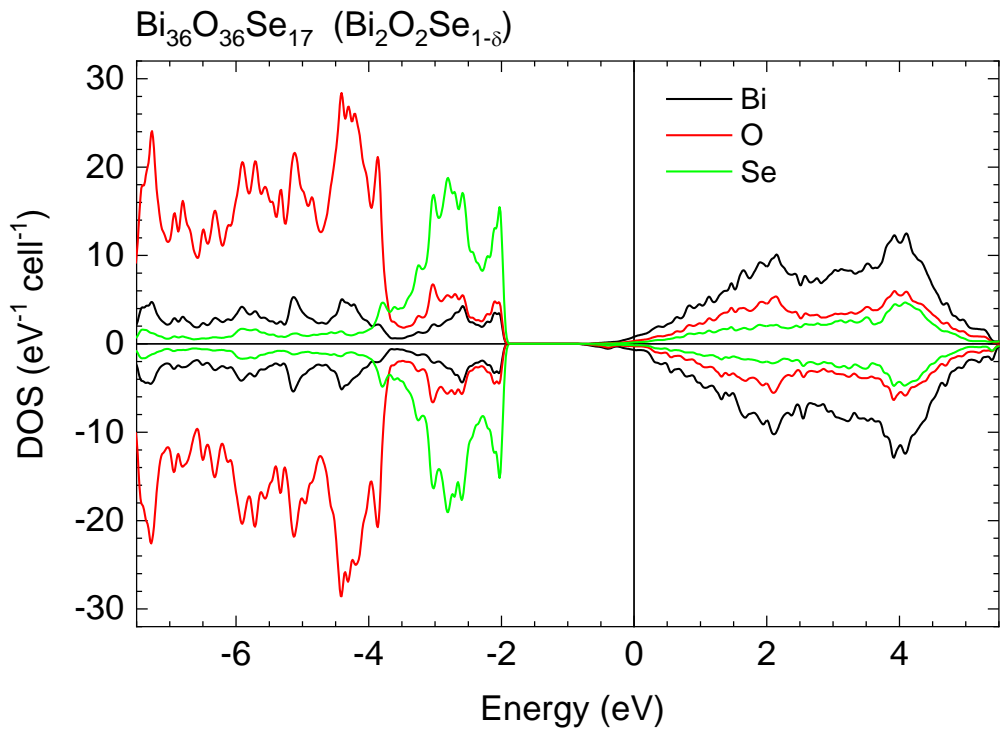

FIG. S9. Partial DOS using mBJ potential for $Bi_2O_2Se$ with Se vacancy $V_{Se}$. The Fermi energy (black vertical line) crosses the conduction band. which indicates n-type metallic behavior.

## 3. IR – reflectivity and transmittance experiments

Figure S10 shows the reflectivity R and transmission T measurement in a broad spectral range. Note a relatively high transmission ≈40%, which spans over large range of frequencies ≈ 1000 – 8000 cm$^{-1}$. The sum R+T[1] ≠1 due to absorption. Four distinct oscillators are seen from the loss function in Figure S11.

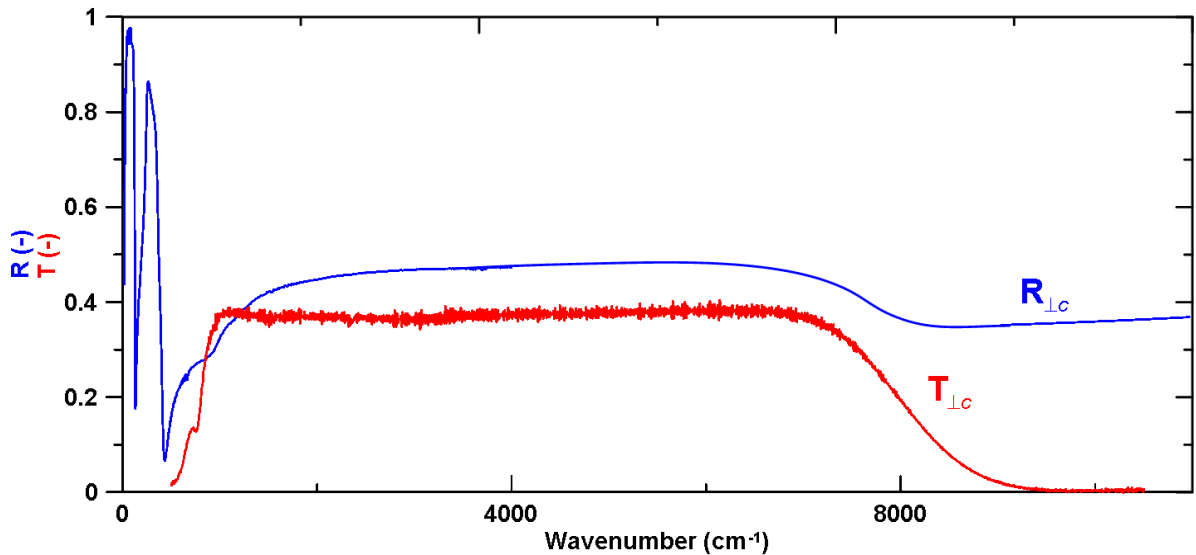

FIG. S10. Broad-range measurement of reflectivity and transmission of $Bi_2O_2Se$ at T=300 K. Sample thickness 0.34 mm. The reflectivity spectrum varies only slightly from sample to sample.

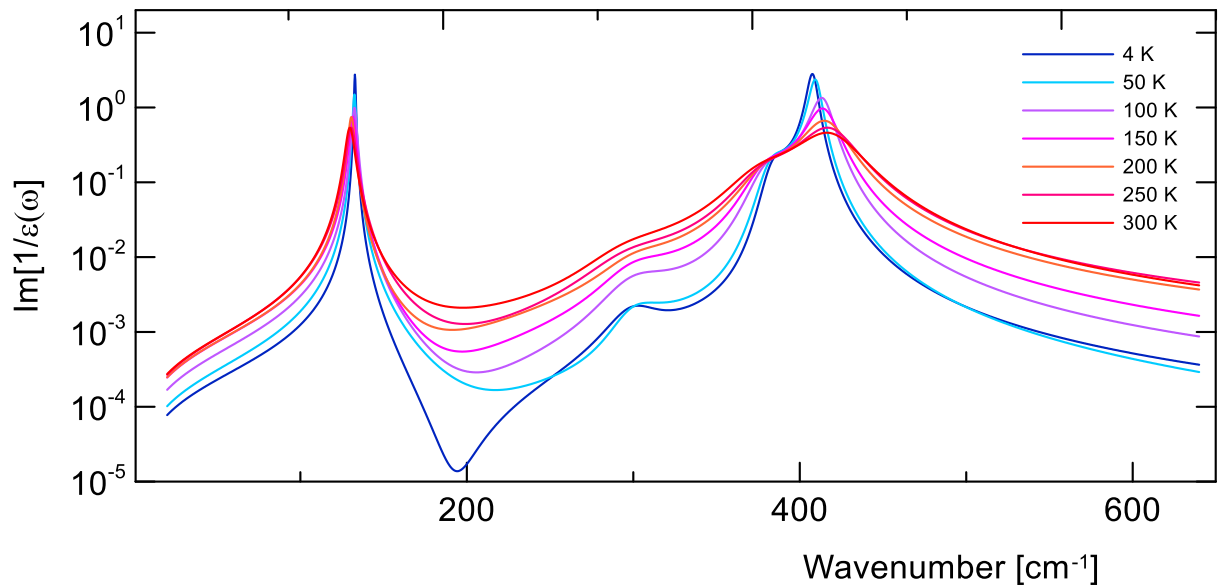

FIG. S11. Peaks of the loss – function $1/\varepsilon_2(\omega)$ indicates frequencies of 4 longitudinal optic modes in the far IR spectral range.

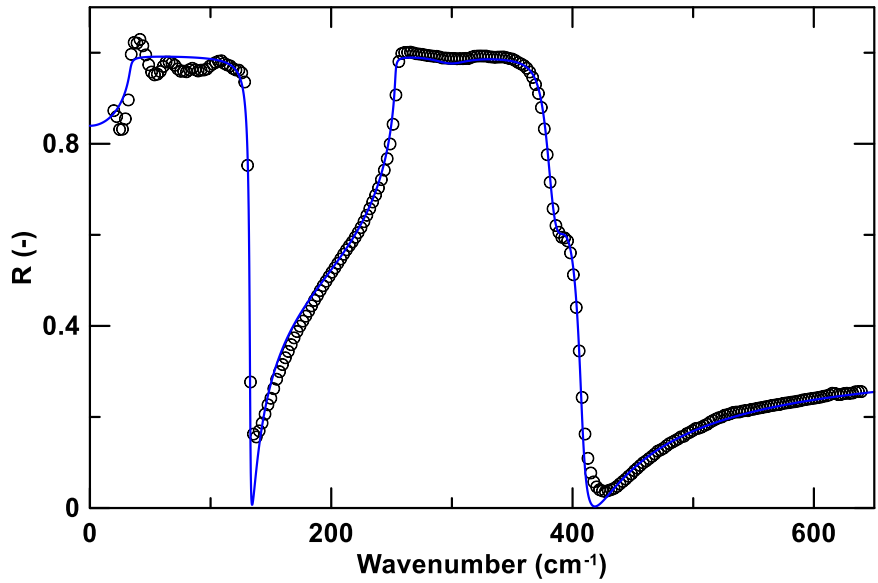

FIG. S12. Near normal reflectivity of $Bi_2O_2Se$ single crystal at 4 K. Empty circles represents experimental data and the blue line is its fit according to the four-parameter model. The details on the model as well as all its parameters are given in the main text.

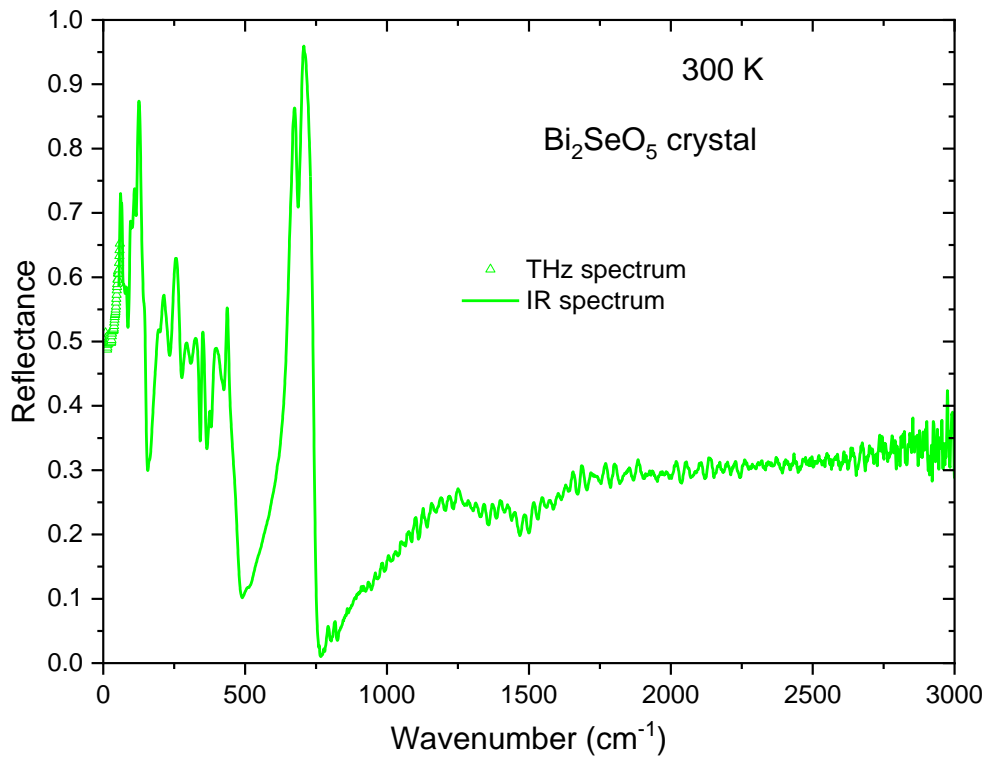

FIG. S13. IR reflectance spectrum of $Bi_2SeO_5$ single crystal. The crystal was 200 μm thin, therefore interferences are seen above 1000 cm$^{-1}$. Symbols below 55 cm$^{-1}$ are data calculated from complex permittivity obtained using time-domain THz transmission spectroscopy. Static permittivity of $Bi_2SeO_5$ obtained from THz spectra is only ~ 30. Note that $Bi_2O_2Se$ single crystal has in-plane permittivity ~ 500. If the $Bi_2SeO_5$ phase were present on the surface of a $Bi_2O_2Se$ crystal, some of the strongest phonons of the $Bi_2SeO_5$ phase would have to be visible in the reflectivity of $Bi_2O_2Se$ crystal. Since we do not see them, we can argue that the weak modes near 300 and 400 cm$^{-1}$ have a multiphonon origin (the sum of two phonon frequencies with the same wave vectors, not necessarily from the center of the Brillouin zone).

## 4. Raman spectra

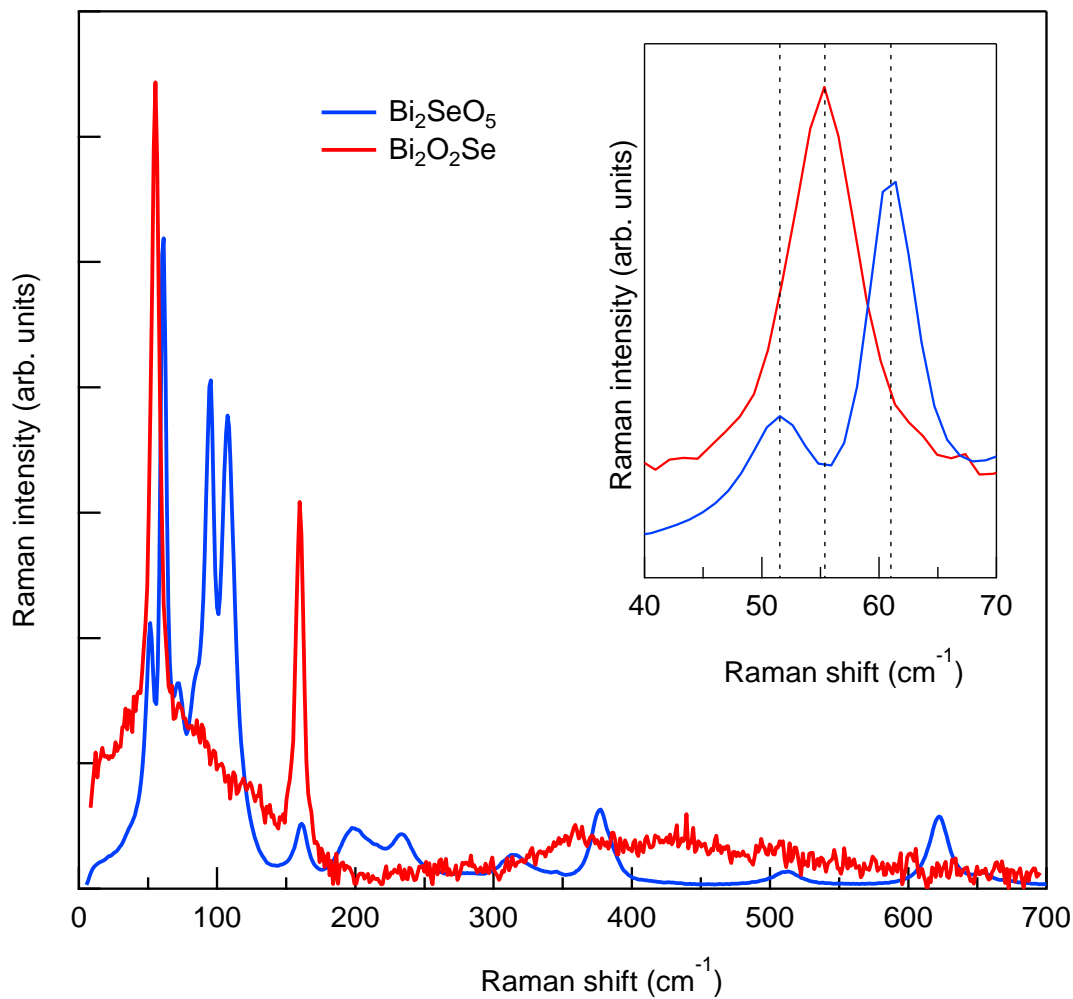

FIG. S14. Comparison of room-temperature Raman spectra of $Bi_2O_2Se$ (red line) and $Bi_2SeO_5$ (blue line) crystals. The enlarged region of the spectra is shown in the inset of the graph in order to show that peak at 55 cm$^{-1}$ correspond to $Bi_2O_2Se$ and the peaks at 51.5 and 61 cm$^{-1}$ are phonons in $Bi_2SeO_5$. Some previous authors who reported the phonon above 60 cm$^{-1}$ observed probably the phonon from secondary $Bi_2SeO_5$ phase.

We have identified all four theoretically predicted Raman modes in the main phase (see Fig. 7 in the main text). No other phonons were detected in the Raman spectra taken on large crystal surfaces, but other phonons corresponding to the $Bi_2SeO_5$ phase were detected at the edges of the sample. The phonons of this phase were also measured on a specially prepared $Bi_2SeO_5$ crystal.
HRXRD of $Bi_2O_2Se$ sample shown extraordinary reflection at Θ ≈18° that is attributable to $Bi_2SeO_5$. This is consistent with the observation of $Bi_2SeO_5$ at the edges of samples in the Raman spectra (see the main text).

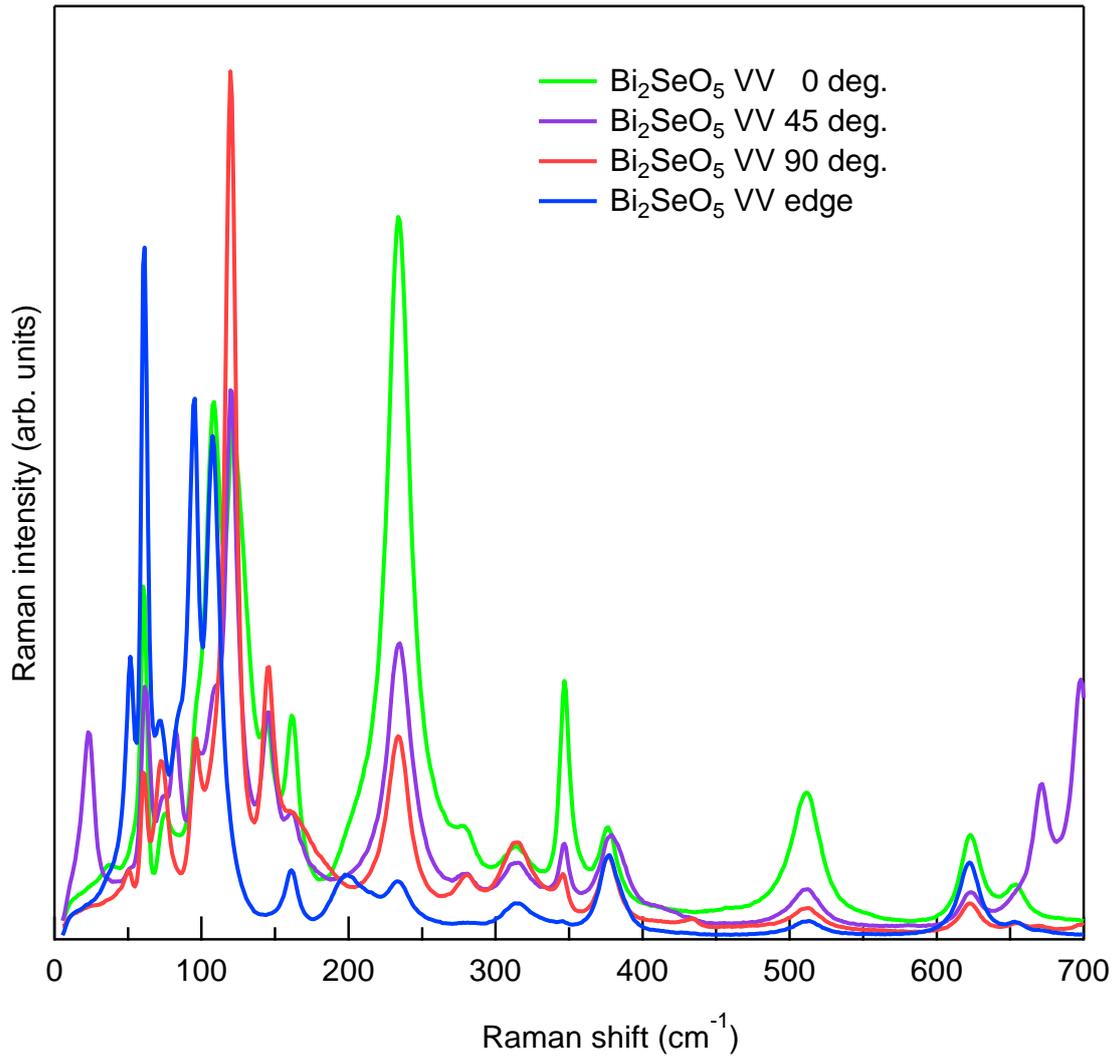

FIG. S15. Room temperature polarized Raman spectra of Bi$_2$SeO$_5$ single crystal taken for different orientations of the sample. VV means that polarizator and analyzator were parallel. Angles mean rotation of the crystal. One spectrum was obtained from the edge of the sample with polarization of incident and scattered light along *c* axis of the crystal.

## 5. DFT calculation – phonon spectra - details

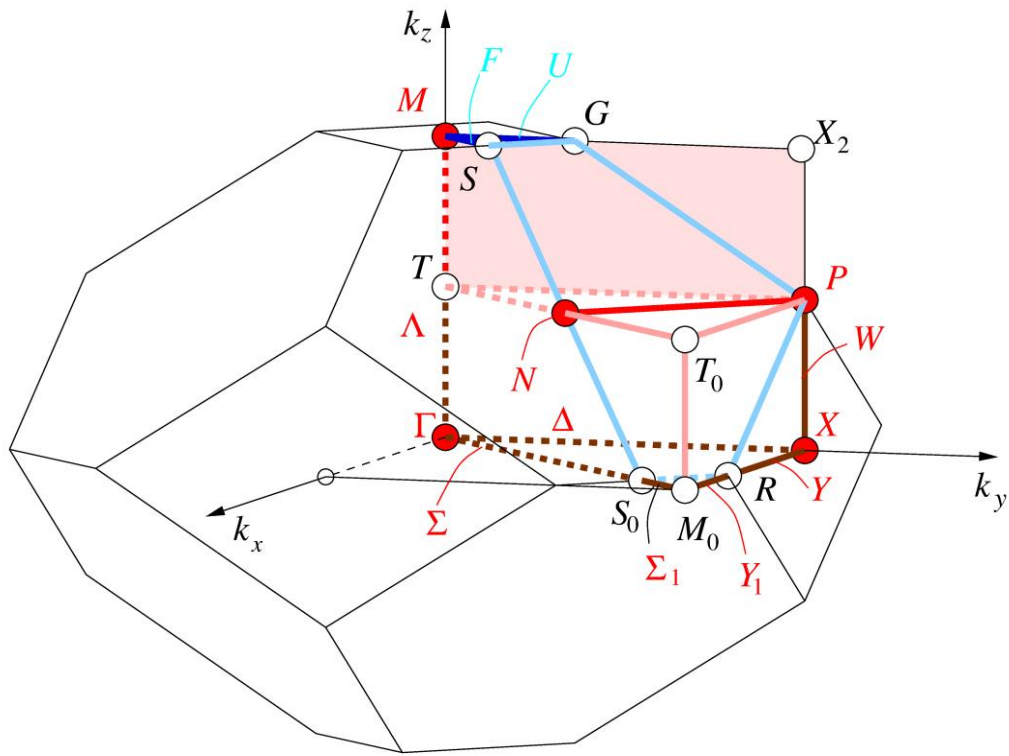

FIG. S16. Brillouin zone of space group I4/mmm with special k-vector points of the asymmetric unit indicated (adapted from https://www.cryst.ehu.es).

Table S2. Electronic contribution to the permittivity $\varepsilon_{el}$ (corresponding to high-frequency permittivity $\varepsilon_\infty$). Ionic contribution to the permittivity $\varepsilon_{ion}$ (which corresponds to the difference between the static and high-frequency permittivity $\varepsilon_r - \varepsilon_\infty$) of $Bi_2O_2Se$, and Born effective charges Z* [9] of the atoms in $Bi_2O_2Se$, calculated using the VASP code. The diagonal elements of *xx, yy,* and *zz* of the 3×3 tensors are shown.

| Tensors | xx, yy | zz |
|---|---|---|
| $\varepsilon_{el}$ | 13.86 | 10.36 |
| $\varepsilon_{ion}$ by DFPT method | 613.9 | 121.3 |
| $\varepsilon_{ion}$ by FDA method | 599.7 | 126.9 |
| Z*(Bi) | 6.12 | 5.40 |
| Z*(O) | -3.96 | -3.83 |
| Z*(Se) | -4.33 | -3.13 |

## 6. Heat capacity and thermal conductivity models

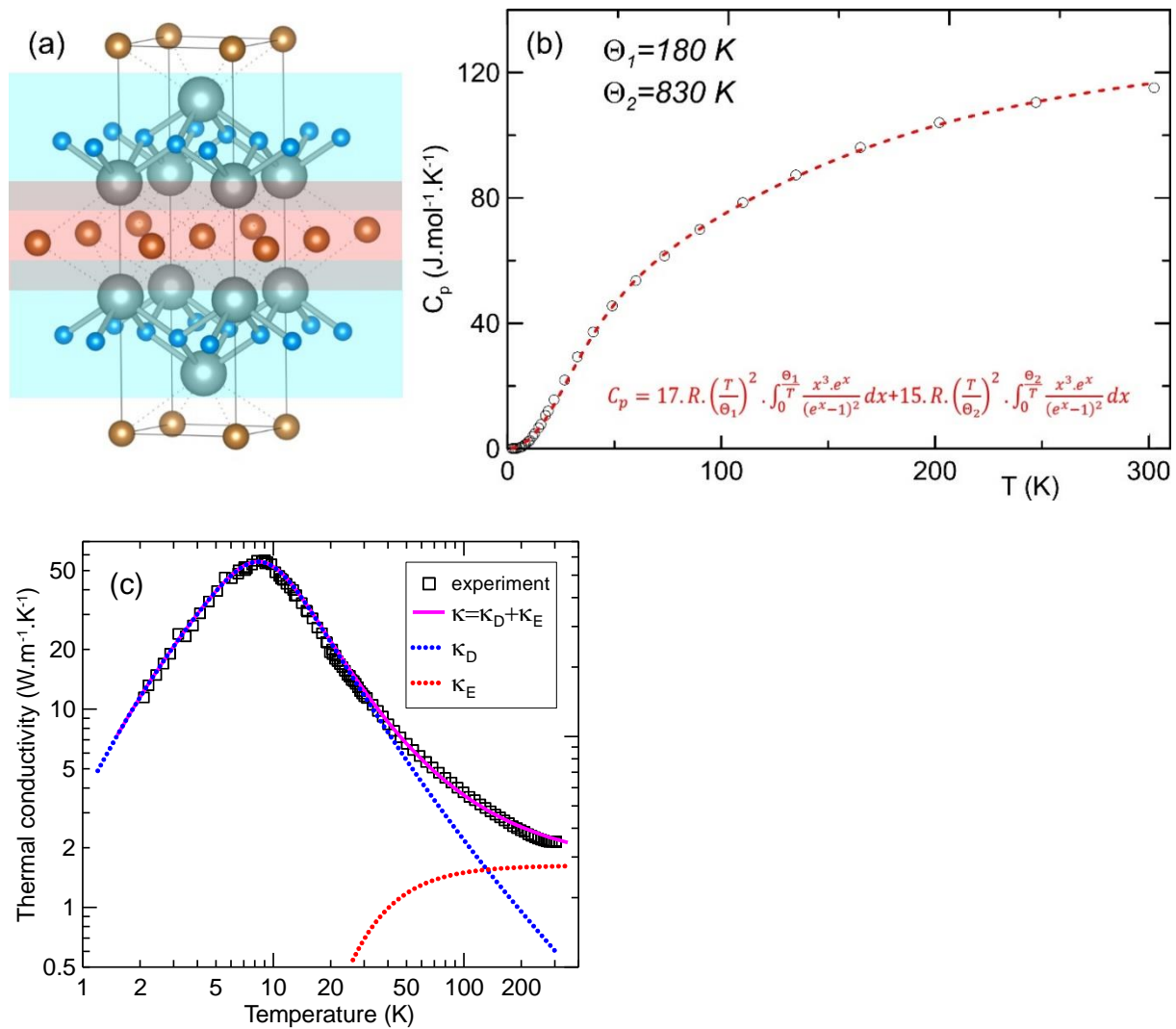

FIG. S17. (a) The $Bi_2O_2Se$ structure can be viewed as a composite of two different 2D structures, one soft (Bi-Se-Bi in red) and one hard (Bi-O-Bi in blue). They have Bi atoms in common. (b) Thus, the heat capacity can be fitted reasonably well using two 2D Debye components, one with a low Debye temperature ($\Theta_1$ =180K) and the other one with a high Debye temperature ($\Theta_2$ =830K). These values are consistent with the two-component model of [10]. In fact, the heat transfer between the layers is limited due to a very different elastic constants, which can be translated as acoustic refractive indices (see the group velocity calculation below). Thus, the phonons can be easily reflected at the interfaces, including total reflection in the Bi-O-Bi to Bi-Se-Bi direction. (c) Theoretical fit (magenta solid line) to the experimental thermal conductivity (symbols). In this case, a Debye model is combined with an Einstein model. The first is designed to account for low energy, largely acoustic phonons and the second for high energy, largely optical phonons. Although the calculated curves are in good agreement with the experiment, a deeper understanding of the temperature dependent thermal conductivity is still lacking. In particular, a connection between the 2D heterogeneous phonon structure (hard and soft layers), the large phonon gap and the anomalous phonon DOS is missing. The deepest analysis so far is provided by the complicated model in [10], which unfortunately deals with out-of-plane thermal conductivity in a polycrystalline material.

### Average group velocity calculation:

We investigated the average group velocity of the phonon contribution. The three components of group velocity $v_g$ at each frequency $\omega_{n,q}$ is obtained from the slope of the dispersion relation:

$$(v_g^{x,y,z})_{n,q} = \frac{\partial \omega_{n,q}}{\partial q_{x,y,z}},$$

where $n$ is the band index in the phonon structure and $q$ is the wavevector. To account for the presence of phonon modes with opposite group velocities in the Brillouin zone, e.g., $(v_g)_{n,q}$ and $(v_g)_{n,-q}$, which have the same magnitude but opposite signs, we consider the root-mean-square (RMS) speed $v_{rms} = \sqrt{\langle v^2 \rangle}$. The contributions from the low-frequency phonons of Bi/Se and the high-frequency phonons of O can be separated as

$$v_{rms}^2(T) = \frac{\sum_{n,q}(v_g)_{n,q}^2 f_{BE}}{\sum_{n,q} f_{BE}} = \frac{\sum_{n=1}^{9}\sum_q (v_g)_{n,q}^2 f_{BE} + \sum_{n=10}^{15}\sum_q (v_g)_{n,q}^2 f_{BE}}{\sum_{n,q} f_{BE}} = \left(v_{rms}^{Bi/Se}(T)\right)^2 + \left(v_{rms}^{O}(T)\right)^2,$$

where $(v_g)_{n,q}^2 = (v_g^x)_{n,q}^2 + (v_g^y)_{n,q}^2 + (v_g^z)_{n,q}^2$ and $f_{BE} = [\exp(\hbar\omega_{n,q}/k_B T) - 1]^{-1}$ is the Bose-Einstein distribution function. In the low-T limit, we obtain

$$v_{rms}(T \to 0) = 2245 \text{ m/s}.$$

In the high-T limit, the values are given by

$$v_{rms}(T \to \infty) = 1530 \text{ m/s},$$

$$v_{rms}^{Bi/Se}(T \to \infty) = 1151 \text{ m/s},$$

$$v_{rms}^{O}(T \to \infty) = 1008 \text{ m/s}.$$

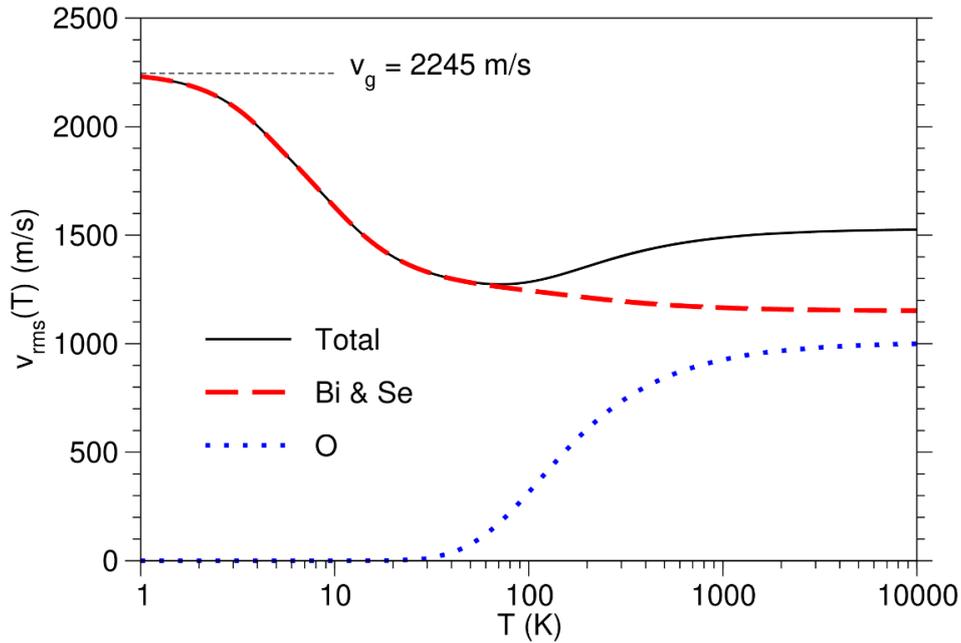

FIG. S18. Calculated $v_{rms}(T)$ from the phonon structure of Bi$_2$O$_2$Se. The total contribution (black solid), Bi and Se phonon modes (red dashed), and O phonon modes (blue dotted) are presented. The same figure is presented in the main text.

## 7. Hall coefficient, electrical conductivity, optical gap

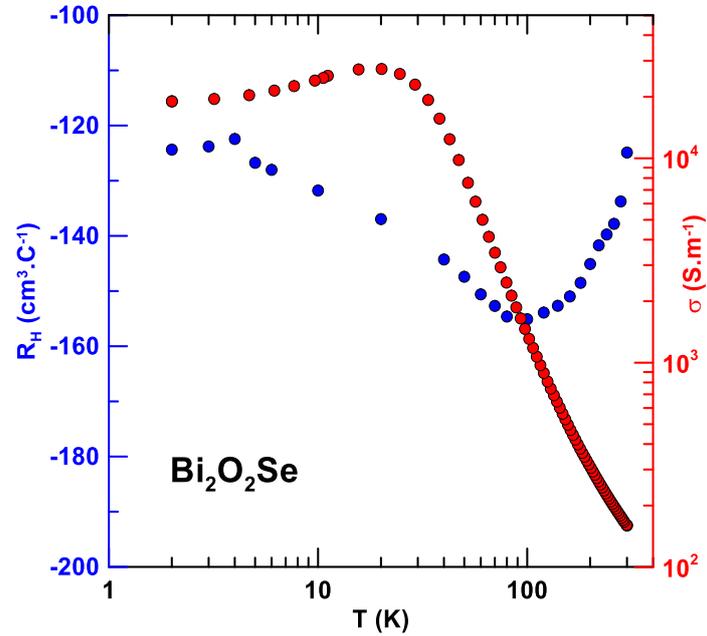

FIG. S19. Hall coefficient and electrical conductivity as a function of temperature for the sample under investigation.

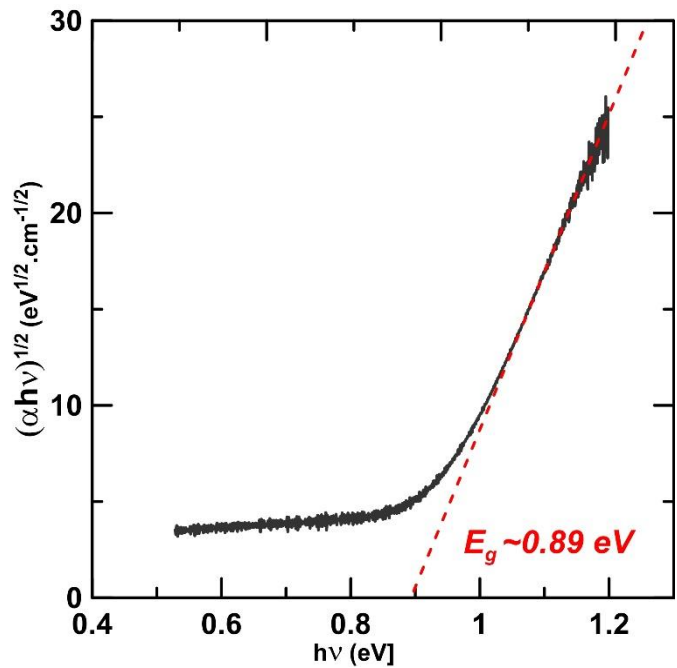

FIG. S20. Optical energy gap $E_{G(opt)}$ of undoped Bi$_2$O$_2$Se evaluated from Tauc's plot for indirect transition. The absorption coefficient $\alpha$ was calculated from the data (R, T) shown in Figure S9 using the formula $\alpha = \frac{1}{d} \ln \frac{(1-R)^2 + \sqrt{(1-R)^2 + 4R^2T^2}}{2T}$.

## 8. Crystal growth – why Se-rich conditions

Here we add some details and comments to the crystal growth of Se-rich crystals presented in the main text in section A. The stoichiometric ratio used for crystal growth naturally leads to Se-rich conditions due to the reaction of part of the $Bi_2O_3$ with the $SiO_2$ ampoule. In fact, it is difficult to create conditions for growing large, flat, Se-poor crystals for optical experiments and the pristine quality of the samples would be destroyed by cutting. In addition to temperature and composition, the temperature gradient plays an important role. In general, both time and temperature are compositional enemies due to the reaction of $Bi_2O_3$ with $SiO_2$.

Note that a higher electron concentration (Se-poor) would be associated with plasma resonance at higher frequencies $\omega_p$ (we do not observe electron plasma resonance), which would prevent any optical analysis of low frequency phonons. We obtain $\omega_p \approx$ 60 cm$^{-1}$ for n=10$^{17}$cm$^{-3}$.

We have also not been able to measure SdH oscillation to better characterize our "optical" sample. The measurement of the SdH effect in the Se-rich samples is difficult to impossible. The Se-poor samples (n>10$^{17}$cm$^{-3}$, residual resistance ratio (R(300K)/R(2K)), RRR>150 samples show nice oscillations (see [11]) in contrast to the Se-rich samples; the present sample´s RRR<120. The electron concentration and mobility and the RRR can be used as an indicator for the Se-rich/Se-poor composition (e.g. [10.11]).